%% file: text-sub.tex
%%%%%%%%%%%%%%%%%%%%%%%%%%%%%%%%%%%%%%%%%%%%%%%%%%%%%%%%%%%%%%%%%%%%%%%%%
%	  ______________________
%	  **********************
%	  *  REVISION HISTORY  *
%	  **********************
%	  ----------------------
%
%	First draft written by SB.
%
%	96-10-24: Mods started by WJC+YPE
%
%	96-11-20: Mods completed -- draft distributed via e-mail to 
%                 authors.
%

%
%\def\figdir{/hosts/big_scr/usr/basu/bison}   % SB definition
%\def\figdir{figs/feb3}   % MJT definition
\def\figdir{fig}   % JCD definition
\overfullrule=0pt
%equation names
\newcount\equationno      \equationno=0
\newtoks\chapterno \xdef\chapterno{}
\newdimen\tabledimen  \tabledimen=\hsize

\def\eqname#1{\global \advance \equationno by 1 \relax
\xdef#1{{\noexpand{\rm}(\chapterno\number\equationno)}}#1}
%tables
\def\table#1#2{\tabledimen=\hsize \advance\tabledimen by -#1\relax
\divide\tabledimen by 2\relax\vskip 1pt
\moveright\tabledimen\vbox{\tabskip=1em plus 4em minus 0.9em
\halign to #1{#2}} }
%

%
% 
% to generate rules within tables
  % top rule
             % middle rule
       % bottom rule
%
% greater than or order of \ga

%
\input epsf
\input mn.tex

\def\jcd{Christensen-Dalsgaard}
\def\notea #1]{{\bf #1]}}
\def\eg{{e.g.}}
\def\cf{{cf.}}
\def\ie{{i.e.}}
\def\etal{{et al.}}
\def\Fsurf{F_{\rm surf}}
\def\d{{\rm d}}
\def\CJ{{\cal J}}
\def\CK{{\cal K}}
\def\Rsun{{\rm R}_\odot}
%
%---------------------------------------------------
%    Beginning of text
%----------------------------------------------------
%

%\Autonumber

\begintopmatter
%\title{An inversion of the solar interior allowing for solar cycle effects}
\title{Solar internal sound speed as inferred from combined BiSON and LOWL
oscillation frequencies}
\author{Sarbani Basu,$^1$ W. J. Chaplin,$^2$ J. Christensen-Dalsgaard,$^1$
Y. Elsworth,$^2$ G.~R.~Isaak,$^2$ R. New,$^3$ J. Schou,$^4$ 
M. J. Thompson$^5$  and S. Tomczyk$^6$}
\vskip 4pt
\affiliation{$^1$Teoretisk Astrofysik Center,  Danmarks Grundforskningsfond,
Institut for Fysik og
Astronomi, Aarhus Universitet, \hfill\break
DK-8000 Aarhus C, Denmark}
\vskip 4pt
\affiliation{$^2$School of Physics and Space Research, University of
Birmingham, Edgbaston, Birmingham, B15, 2TT, U.K.}
\vskip 4pt
\affiliation{$^3$School of Science and Mathematics, Sheffield Hallam
University, Sheffield, S1 1WB, U.K.}
\vskip 4pt
\affiliation{$^4$Center for Space Science and Astrophysics, HEPL Annex A201,
    Stanford University, Stanford, CA 94305, U.S.A}
\vskip 4pt
\affiliation{$^5$Astronomy Unit, School of Mathematical Sciences, Queen
Mary and Westfield College, London E1 4NS, U.K.}
\vskip 4pt
\affiliation{$^6$High Altitude Observatory, NCAR, P.O. Box 3000, Boulder, 
CO 80307, U.S.A.}

\shortauthor{S. Basu et al.}
\shorttitle{The solar core}
\acceptedline{Accepted \ . Received \ }

\abstract{Observations of the Sun with the LOWL instrument
provide a homogeneous set of solar p-mode
frequencies from low to intermediate degree which allows one to
determine the structure of much of the solar interior
avoiding systematic errors
that are introduced when different data sets are combined, i.e.,
principally the effects of solar cycle changes on the frequencies.
Unfortunately, the LOWL data set contains very few of the
lowest-degree modes, which are essential for determining reliably the
structure of the solar core -- in addition, these lowest-degree
data have fairly
large associated uncertainties. However, observations made by the
Birmingham Solar-Oscillations Network (BiSON) in integrated sunlight
provide high-accuracy measurements of a large number of low-degree
modes.
In this paper we demonstrate that the low-degree mode set of the LOWL
data can be successfully combined with the more accurate BiSON data,
provided the observations are contemporaneous for those frequencies
where the solar-cycle-induced effects are important.  We show that
this leads to a factor-of-two decrease in the error on the inferred
sound speed in the solar core. We find that the solar sound speed is
higher than in solar models for $r < 0.2\Rsun$.  The
density of the solar core is, however, lower than that in solar
models.}


\maketitle
%\Referee

\section{Introduction}

Observations of the Sun, covering many years, have now provided us
with accurate measurements of solar p-mode frequencies which impose
severe constraints on the structure of the Sun.

The LOWL instrument (LOWL is an abbreviation for low degree with 
degree denoted by L) has,
for the first time, provided us with a uniform set of frequencies
from low to intermediate degree $l$. This has allowed detailed inversions
for the structure of much of the Sun's interior (cf., Basu et al.
1995; 1996a; 1996b) to be performed.  The available data from the
instrument now span a period of more than 1 year, and most of the
mode-frequency determinations have relative errors as small as a few
parts in $10^6$.

The LOWL data set contains very few of the lowest-degree modes, i.e.,
for $0 \le l \le 2$, and those that are present have much larger
errors than their higher-$l$ counterparts. Since only the
low-degree modes penetrate to the solar core, large uncertainties in
the inferred structure of the solar core arise from inversions which
rely solely on these data. Observations made by the Birmingham
Solar-Oscillations Network (BiSON) do, however, provide very accurate
measurements of low-degree modes (cf.  Elsworth et al.  1994). These
modes have to be combined with intermediate and high-degree modes
from other sources before they can be used to infer the solar
structure by inverting the observed frequencies -- it would therefore
seem logical to combine the BiSON and LOWL solar data in order to
provide a more reliable determination of the structure of the solar core.

\begintable{1}
\caption{\bf Table 1. \rm Summary of data sets}
\halign{#\hfil &\quad  # \hfil & \quad # \hfil \cr
&Name & Summary details\cr
\noalign{\medskip}
1 &BiSON-8 & 8-month BiSON spectrum; maximum \cr
&&likelihood analysis\cr
\noalign{\smallskip}
2 &BiSON-G & Five 2-month BiSON spectra, Gaussian-\cr
&&analysed\cr
\noalign{\smallskip}
3 & & Lower frequency data from 32-month BiSON\cr
&&spectrum; maximum likelihood analysis\cr
\noalign{\smallskip}
4 & & Comparison BiSON data taken at high solar\cr
&&activity; Gaussian-analysed\cr
\noalign{\smallskip}
5 &LOWL & 1-year spatially-resolved LOWL data\cr
\noalign{\smallskip}
6 &Best Set & Combination of BiSON sets 1-3 and LOWL\cr
}
%\tabletext{}
\endtable

Great care must be taken in combining data from different
sources. Systematic differences 
may arise from differences in instrument characteristics
and analysis techniques, or from temporal variations
in the solar p-modes if the different observations are not
contemporaneous.  Such differences 
may be interpreted by an inversion as being solar in origin, and in
particular in the present context as arising from {\it spatial}
variations in the solar interior. 

The issue of combining non-contemporaneous data is particularly 
important, because it
is known that the p-mode frequencies change along the solar cycle
(Libbrecht \& Woodard 1990; Elsworth et al. 1994). 
The difficulties in determining the solar core
structure from inhomogeneous data sets have been discussed by Gough
\& Kosovichev (1993), Gough, Kosovichev \& Toutain (1995) and
Basu {\etal} (1995, 1996a). 
The solar-cycle variations
are believed to be the result primarily
of variations in the surface properties
of the Sun.  It may therefore be possible that the temporal variations
can be removed in the inversion
in the same way as
other surface uncertainties ({\cf} Dziembowski et al. 1991;
Kosovichev et al. 1992) but, as was pointed out 
by Basu {\etal} (1996a), there remain significant
problems in the combination procedure
and the resulting inversions may be misleading.

We have largely avoided such problems in the present work by 
combining near-contemporaneous data from the BiSON network and the
LOWL instrument, taken at a low-activity phase of the solar
cycle. We also use some low-frequency BiSON data 
based on observations over
several years, but such low-frequency modes will be relatively 
insensitive to solar-cycle variations in the near-surface layers
({\cf} Libbrecht \& Woodard 1990). We also present results of 
combining LOWL with BiSON data collected at
times of high solar activity, to illustrate the dangers inherent
in inverting non-contemporaneous data.

\section{The data}

The data used for the analysis come from two sources: the LOWL
instrument based on Mauna Loa (Tomczyk {\etal} 1995) in the Hawaiian Islands; and the
global, 6-station BiSON network (Chaplin et al. 1996a).  
Both systems employ the same basic physical principles -- the use of
an atomic standard -- to measure the Doppler velocity shift of a
solar Fraunhofer line formed by potassium atoms in the near infrared
($770\,\rm nm$).  The LOWL instrument spatially resolves the visible
solar disc, and is therefore sensitive to oscillation modes of degree
up to $l = 99$.  The BiSON instruments view the unresolved Sun,
and are sensitive to modes of $0 \le l \le 4$.  Although limited
to the lowest-$l$ modes, the BiSON technique is very stable and
provides some of the highest-quality measures of those modes
available.  The combination of these two data sets should be very
powerful given that: they result from observations made on the same
line in the solar atmosphere; each are characterized by high
intrinsic accuracies; and both were collected at the same epoch.

The LOWL time series used for the analysis covers a 12 month period
beginning 1994 February 26 and has a temporal duty cycle of
22 per cent.  Further details of the LOWL experiment are given by
Tomczyk {\etal} (1995).
The data were analysed up to $3.5\,\rm mHz$, which thus set the
upper limit to the frequency range used in this paper, while the
lower limit was set by the visibility of the modes in the data set. 
The fitting procedure, which is a maximum-likelihood technique
applied to the complex Fourier transform of the time series,
is described in Appendix~5 of Schou (1992).

\beginfigure*{1}
%\vskip 7.5 true cm
\hskip  .5 true cm
\vbox to 7.5 true cm{\vskip 0 true cm
\epsfysize=7.5 true cm\epsfbox{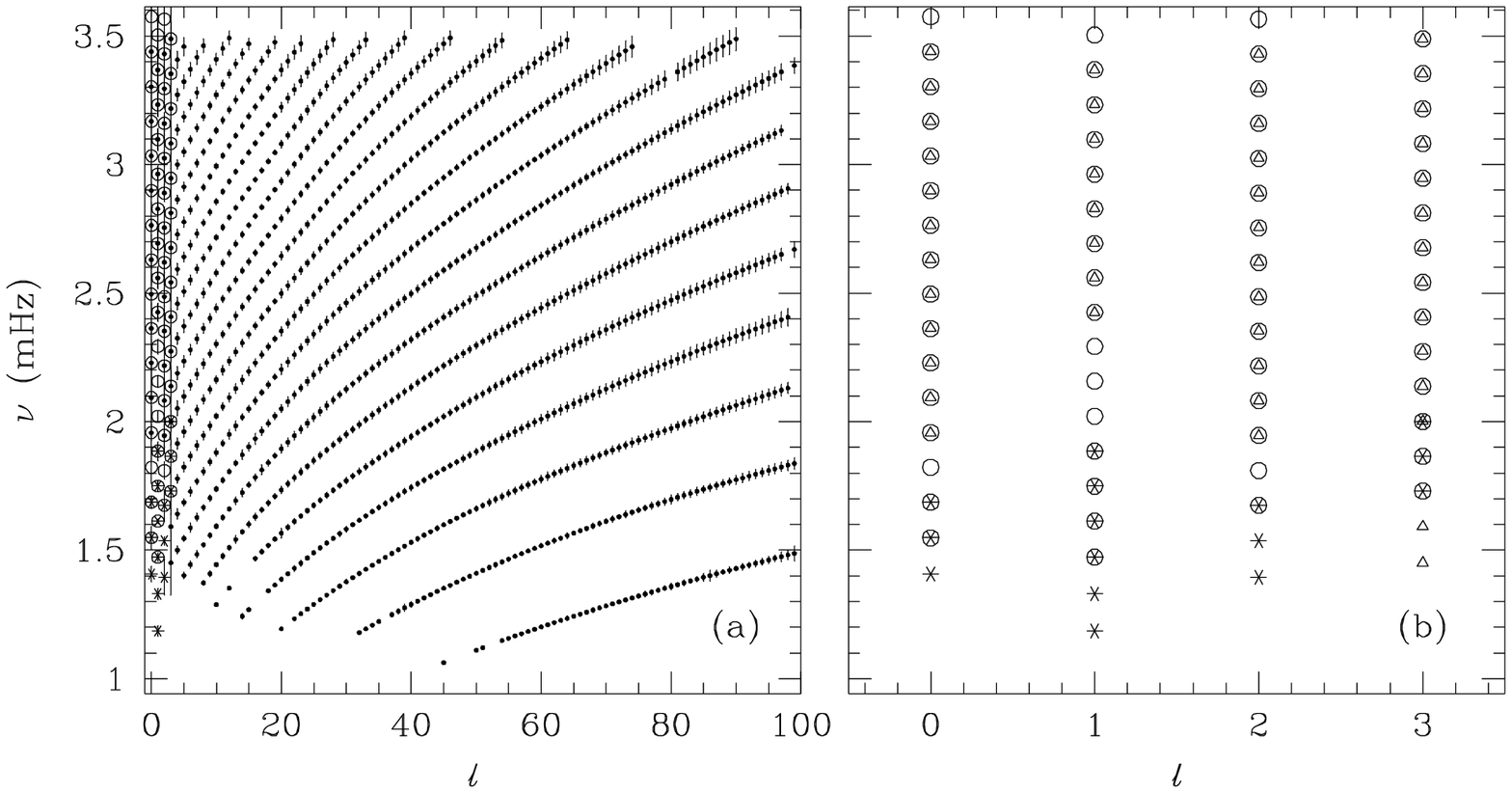}\vskip 0 true cm}
\caption{\bf Figure 1. \rm The $l$-$\nu$ diagram of the data used.
Panel (a)  shows the entire range with 1000$\sigma$ error bars, while
 panel (b) shows
just the low-degree modes.  The triangles (dots in panel a) are LOWL modes; 
circles are
BiSON-8 modes; asterisks are 32-month BiSON modes.}
\endfigure

Several BiSON data sets have been used: these are summarised briefly
in Table~1.
For reasons of convenience, we have used an
8-month BiSON spectrum within the time span of the LOWL set
(henceforth referred to as BiSON-8), generated from data collected
between 1994 January 26 and 1994 September 22. The duty cycle of the
BiSON-8 time series was 80 per cent. 
The modes in the Fourier spectra generated from the BiSON-8
time series were fitted by minimizing a
maximum-likelihood function that assumed an underlying $\chi^2$
distribution with two degrees of freedom
in the frequency domain
(e.g., Anderson, Duvall \& Jefferies 1990; Toutain \& Appourchaux 1994).

In addition, we have fitted
frequencies from five 2-month BiSON spectra (which are
contemporaneous with the LOWL data).
The duty cycle for the complete calendar year 1994 was 78 per cent.
The modes in the
2-month BiSON spectra were first smoothed before fitting to yield
pseudo-Gaussian statistics in the frequency domain -- these data are
therefore referred to as the BiSON-G set.

To get reliable
determinations of the lowest-degree modes at frequencies below
$1.8\,\rm mHz$ one requires spectra generated from more than 1 year's
data, and for this reason we used results derived from the
not-entirely contemporaneous 32-month BiSON spectrum.  These modes
are not expected to be influenced significantly by the solar cycle,
and we do in fact demonstrate that their inclusion in the analysis
does not bias the results.  
Mode frequencies were obtained from a single, 32-month BiSON
spectrum of observations between 1992
October and 1995 June (Chaplin et al. 1996b).  The 
fractional fill of useful data was 72 per cent.
The spectrum was analysed in the same way as the 8-month spectrum.

Finally, to show that variations in p-mode frequencies over the 
solar-cycle do have a deleterious effect on structural inversions,
we have used a set of frequencies derived from BiSON observations
made at times of high solar activity. These data are described by 
Elsworth {\etal} (1994).
 
The fitted BiSON frequencies at $l=0$ and 1 were, in general, more
accurate than their LOWL counterparts; for $l=2$, the quality of
the data were similar; while for $l=3$, the LOWL frequencies were,
by and large, better determined.  We have derived a ``Best Set'' of
frequencies by merging the available data, giving due and careful
consideration to the comparative quality of the frequency
determinations in each set (see later).  The LOWL and BiSON data sets
are plotted in Fig.~1 in the form of an $l$-$\nu$ diagram.  Because the
lower turning points of the modes in the combined data set span the 
range of most radii from the Sun's surface to its centre, and because the
errors on the frequencies are very small, it is
feasible to perform a meaningful
inversion for the solar structure throughout most of the interior.

\section{Inversion Techniques and Solar Models}

Solar oscillations can be described
throughout most of the solar interior by equations describing linear
adiabatic oscillations (cf. Unno et al. 1989). Chandrasekhar (1964)
showed that these equations, along with appropriate boundary
conditions, constitute a self-adjoint eigenvalue problem, which leads
to a variational principle connecting the eigenfrequencies to the
basic equilibrium state of the Sun. 

Our inversion for solar structure 
is based on linearizing
the equations of stellar oscillation around a known reference
model
({\cf}, Dziembowski, Pamyatnykh \&
Sienkiewicz 1990; D\"appen {\etal} 1991; Antia \& Basu 1994;
Dziembowski {\etal}~1994).
The differences between the structure of the Sun and the
reference model are then related to the differences in the
frequencies of the Sun and the model by kernels.

Non-adiabatic effects give rise to frequency shifts (Cox \& Kidman
1984; Balmforth 1992) which are not accounted for by the variational
principle.  Frequency shifts are also introduced by errors in
modelling the underlying solar model, e.g., the effects of turbulent
convection.  Most of these uncertainties affect the surface layers of
the models.  In the absence of any reliable formulation, these
effects have been taken into account in an {\it ad hoc} manner by
including an arbitrary function of frequency in the variational
formulation (Dziembowski et al. 1990). This can be justified because
in the surface layers, the eigenfunction is largely independent of
the degree $l$
of the mode; thus, for spherically symmetric
perturbations, the frequency difference resulting from the surface
effects, weighted by the inertia of the mode, is roughly a function
of frequency only ({\jcd} \& Berthomieu 1991).  We thus express the
fractional change in frequency of a mode in terms of fractional
changes in the model parameters and also a surface effect.

When the oscillation equation is linearised -- under the assumption
of hydrostatic equilibrium -- the fractional change in the
frequency can be related to the fractional changes in two of the
model parameters.  Thus, 
$$
\eqalign{ {\delta \omega_i \over \omega_i}& 
= \int K_{1,2}^i(r){ \delta f_{1}(r) \over f_1(r)}\d r +
 \int K_{2,1}^i(r) {\delta f_{2}(r)\over f_2(r)} \d r \cr &\quad
 +{\Fsurf(\omega_i)\over Q_i} \; \cr}\eqno\eqname\full 
$$ 
({\cf} Dziembowski {\etal} 1990).  Here $\delta \omega_i$ is the
difference in the frequency $\omega_i$ of the $i$th mode between the
solar data and a reference model.  The functions $f_{1}$ and $f_{2}$
are an appropriate pair of model parameters.  The kernels $K_{1,2}^i$
and $K_{2,1}^i$ are known functions of the reference model which
relate the changes in frequency to the changes in $f_{1}$ and $f_{2}$
respectively; and $Q_i$ is essentially the inertia of the mode 
(Christensen-Dalsgaard 1986).  The term in $\Fsurf$
results from the near-surface errors.

The pair of variables $(f_1, f_2)$ can involve several combinations of model
parameters. As discussed below, our goal in the inversion is to 
isolate one of the variables while endeavouring to ensure that 
our results are insensitive to the other variable.
In this work we generally use $(c^2,\rho)$,
$c$~being adiabatic sound speed and $\rho$ density,
to invert for $c^2$.
To invert for density, one good pair to use is $(\rho,\Gamma_1)$,
where $\Gamma_1$ is the first adiabatic exponent. 
Alternatively, one may
assume the equation of state to be known and transform the dependence of the
oscillations on $\Gamma_1$ into a dependence instead on $Y$
($Y$ being the abundance by mass of helium), 
to carry out inversion for the pair $(\rho, Y)$.
(As formulated, this also assumes that the heavy element abundances are known.) 
We have used this formulation,
which has the advantage that
the frequency dependence on $Y$ is largely confined to
the helium ionization zones. This is easier to suppress in the inversion
for density than the dependence on $\Gamma_1$ would be in an inversion using
$(\rho,\Gamma_1)$. Consequently we can achieve better resolution and/or
error properties in our density inversion. However, this advantage is bought
by assuming the equation of state to be known, 
a point to which we return in Section~5.

A number of different inversion techniques can be used for inverting
the constraints given in equation~\full. We have used the
subtractive optimally localised averages (SOLA) method of Pijpers \&
Thompson (1992).

\begintable{2}
\caption{\bf Table 2. \rm Properties of solar models}
\halign{#\hfil &  \hfil # \hfil &
\hfil # \hfil & \hfil # \hfil & \hfil # \hfil & \hfil # \hfil &
\hfil # \hfil & \hfil # \hfil & \hfil # \hfil \cr
Model&$(Z/X)_{\rm s}$ &$Y_{\rm e}$ & $Y_{\rm c}$ &$T_{\rm c}$&
$\rho_{\rm c}$& $r_{\rm d}/\Rsun$ & Age\cr
 & & & &$10^6$ K& g cm$^{-3}$ & & Gyr \cr
\noalign{\medskip}
Reference  & 0.0245 & 0.2447 & 0.6444 & 15.67 & 154.2 & 0.7115 & 4.6\cr
Test  &0.0245  & 0.2457  & 0.6402 & 15.64 & 152.9 & 0.7124 & 4.52\cr
}
\tabletext{Notes: $(Z/X)_{\rm s}$ is the present  surface heavy element
 abundance 
ratio; $Y_{\rm e}$ and $Y_{\rm c}$ are  the current envelope and central helium
 abundances respectively; $T_c$ is the central temperature, $\rho_c$ the
central density, and $r_{\rm d}/\Rsun$ the radial location of the base of
 the convective envelope in the models used in this paper.}
\endtable

The principle of the inversion technique is to form linear
combinations of equations (1) with weights $c_i(r_0)$ chosen such as
to obtain an average of $\delta f_1/f_1$ localized near $r = r_0$
while suppressing the contributions from $\delta f_2/f_2$ and the
near-surface errors.  In addition, the statistical errors in the
combination must be constrained.  If successful, the result may be
expressed as
 $$
 \int \CK(r_0, r) {\delta f_1(r) \over f_1(r)} \d r
 \simeq \sum c_i(r_0) {\delta \omega_i \over \omega_i} \; ,\eqno\eqname\avker
 $$ 
 where the $\CK(r_0, r)$, the averaging kernel at $r=r_0$, is
defined as
 $$
 \CK(r_0, r) = \sum c_i(r_0) K_{1,2}^i(r)  \; ,
 \eqno\eqname\avdef
 $$ 
 of unit integral, and determines the extent to which we have
achieved a localized measure of $\delta f_1/f_1$.  In particular, the
width in $r$ of $\CK(r_0,r)$, here calculated as the distance between
the first and third quartile point, provides a measure of the
resolution\note{$^\dagger$}{The radii $r_1$, $r_2$, $r_3$
of the first, second and third quartile 
points are defined such that $\CJ(r_1)=0.25$, $\CJ(r_2)= 0.5$ and
$\CJ(r_3)=0.75$, where $\CJ(r) = \int_0^r\CK\d r$ (recall that
$\CK$ is unimodular).
In practice this uniquely defines $r_1$, $r_2$ and $r_3$ for our kernels.}.
The precise implementation of the method is described by
Basu et al.  (1995, 1996c).

A drawback of the SOLA method is the implicit assumption
that the frequency differences and the errors associated with them
are correct. However, sometimes the observational errors can be
either under- or overestimated ({\eg} because the modes have 
much lower line widths than modes adjacent in
frequency), in which case the inversion result can be misleading.  We
have found that a reliable way to detect such modes is to do a
regularized least squares (RLS) inversion first. We rejected modes
which had more than a $3.5\sigma$ residual after the fit. The SOLA
inversion was performed on the weeded mode set. On examining some of
the low-degree modes that had been rejected by the RLS procedure, we
found that the fit to the mode concerned was usually poor.  Thus the
RLS is a reasonably reliable way of weeding out uncertain modes.
There are some interesting instances where, even in contemporaneous
data, the appearance of a particular mode was very different in each
data set, presumably because of beating with noise. In addition,
different analysis techniques can give rise to fitted frequencies
which differ by more than the formal error. The details of the RLS
implementation can be found in Antia \& Basu (1994).

Our inversion requires the use of a reference solar model.
The model we use (Model S of {\jcd} et al. 1996) was constructed with
the Livermore (OPAL) equation of state (Rogers, Swenson \& Iglesias,  1996).  For
temperatures higher than $10^4$ K, OPAL opacities are used 
(Iglesias, Rogers \& Wilson 1992),
whereas at lower temperatures opacities from the tables
of Kurucz (1991) are taken. The model incorporates the diffusion of
helium and heavy elements below the convection zone. The surface
heavy element ratio is $Z/X = 0.0245$ (Grevesse \& Noels 1993).  The
model has an age of 4.6 Gyr. We have also used another model as a
proxy Sun to illustrate certain effects.  The physical assumptions in
the
model are identical to those in the reference model, but it has a lower age of
4.52 Gyr. Some of the properties of the models are listed in Table~2.

\beginfigure{2}
%\vskip 6.5 true cm
\hbox to 0 pt{\hskip -1.5cm
\vbox to 6.5 true cm{\vskip -1.75 true cm
\epsfysize=10.50 true cm\epsfbox{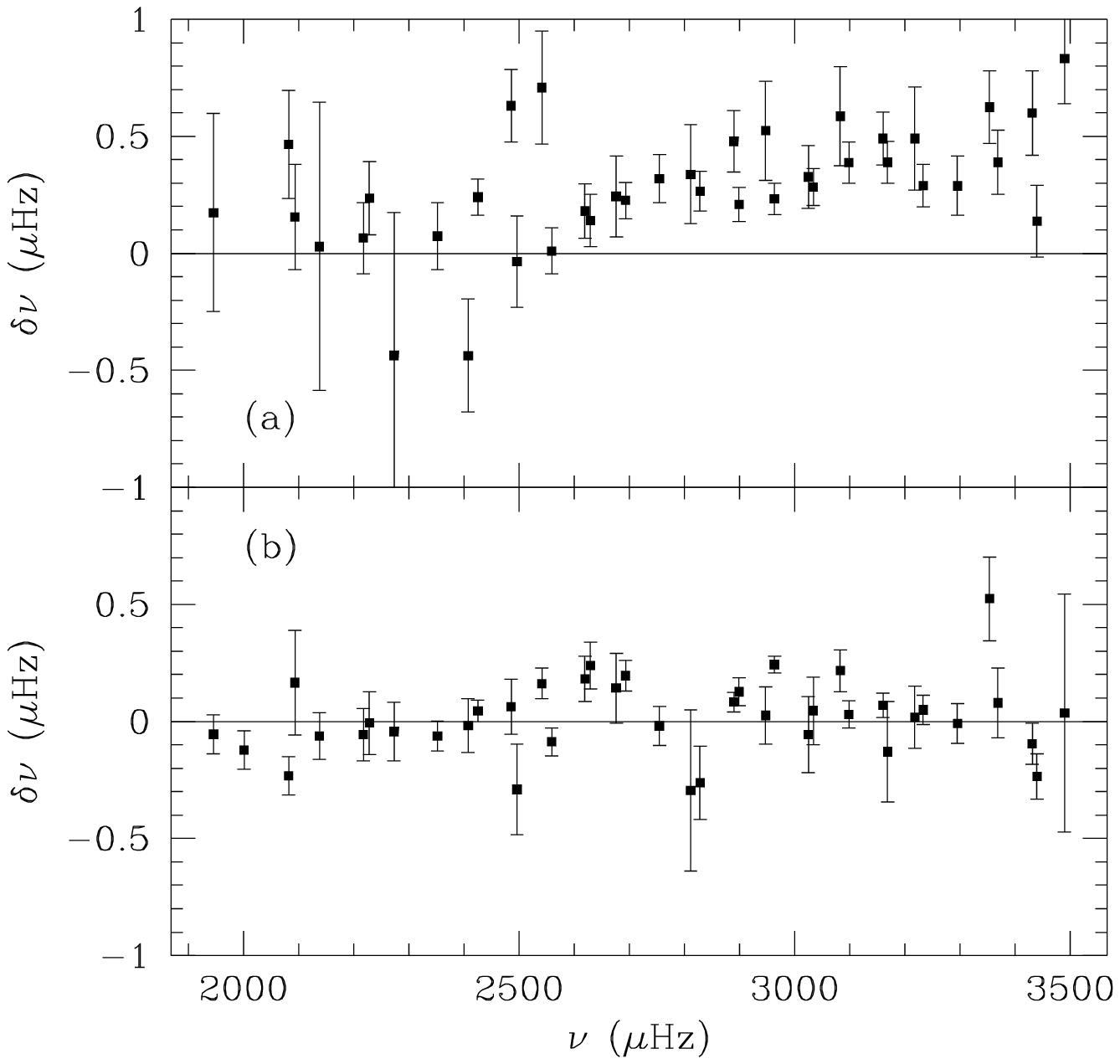}\vskip -2.0 true cm}}

\caption{\bf Figure 2. \rm (a) The frequency differences between the
high-activity BiSON and LOWL data sets.  (b) The frequency
differences between the BiSON-8 and LOWL data sets. The error bars on
each plot were generated by adding the respective BiSON and LOWL
uncertainties in quadrature.}
\endfigure

\section{Results}

\subsection{ Consistency of the frequencies}

To illustrate the effects of the solar cycle on the observed
frequencies, the frequency differences between the LOWL and the
BiSON-8 and BiSON high-activity data 
have been plotted in Fig.~2.  The high-activity BiSON data show
a systematic, frequency-dependent difference with respect to the LOWL
data. The BiSON-8 data show no such trend.  Since slowly varying,
frequency-dependent terms in the differences are caused by
perturbations near the solar surface (e.g. Gough 1990), this implies
that while the BiSON-8 and LOWL data show similar surface effects,
the BiSON high-activity data do not.  The differences between the
high-activity BiSON and LOWL data are indicative of the influence of
the solar cycle. Consequently, if the low-degree LOWL data were to be
replaced by the high-activity BiSON data, the combined mode set would
have two different surface terms.

\beginfigure{3}
%\vskip 8.5 true cm
\hbox to 0 pt{\hskip -2.5cm
\vbox to 16 true cm{\vskip -2.0 true cm
\epsfysize=12.50 true cm\epsfbox{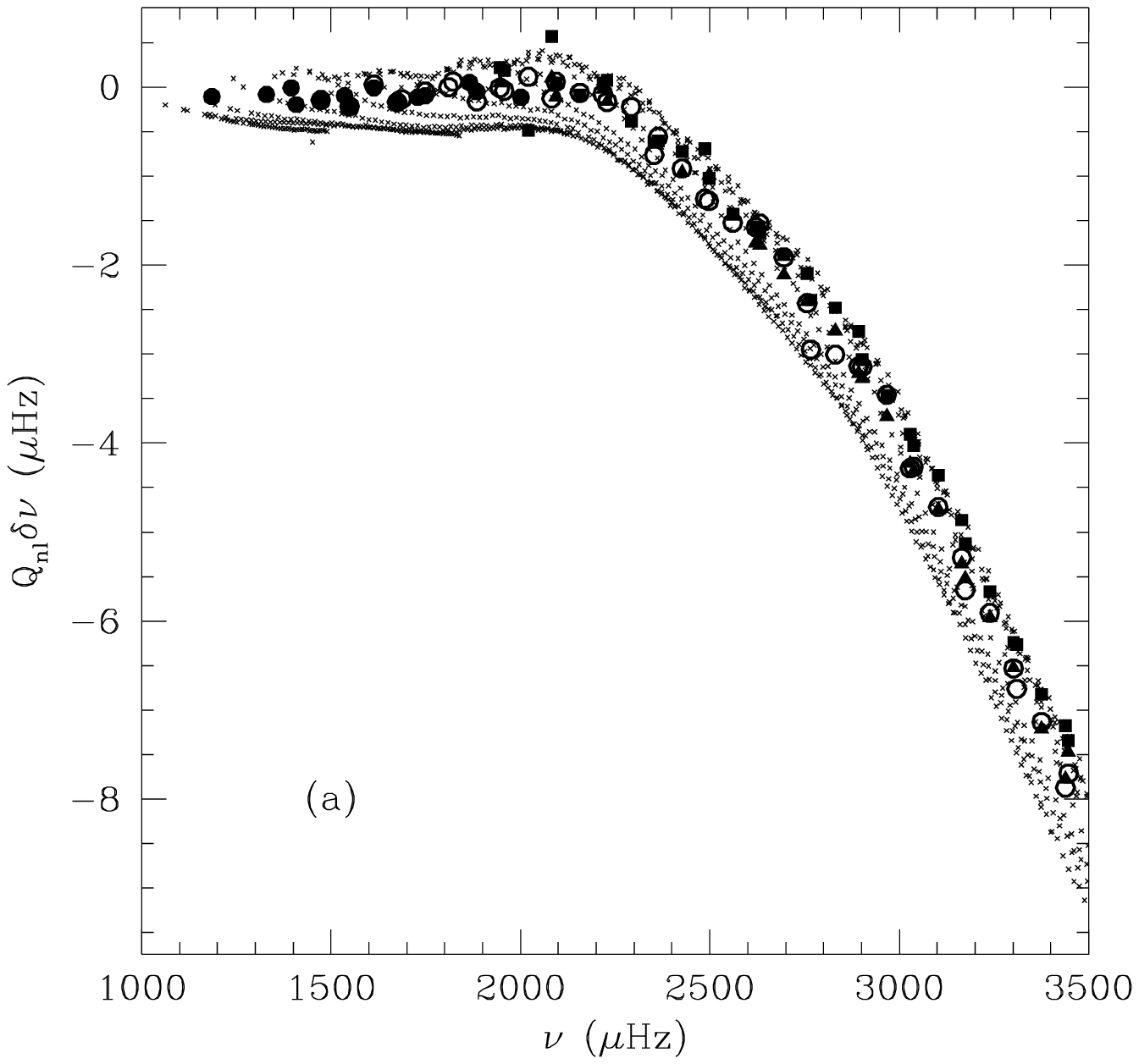}\vskip -4.5 true cm
\epsfysize=12.50 true cm\epsfbox{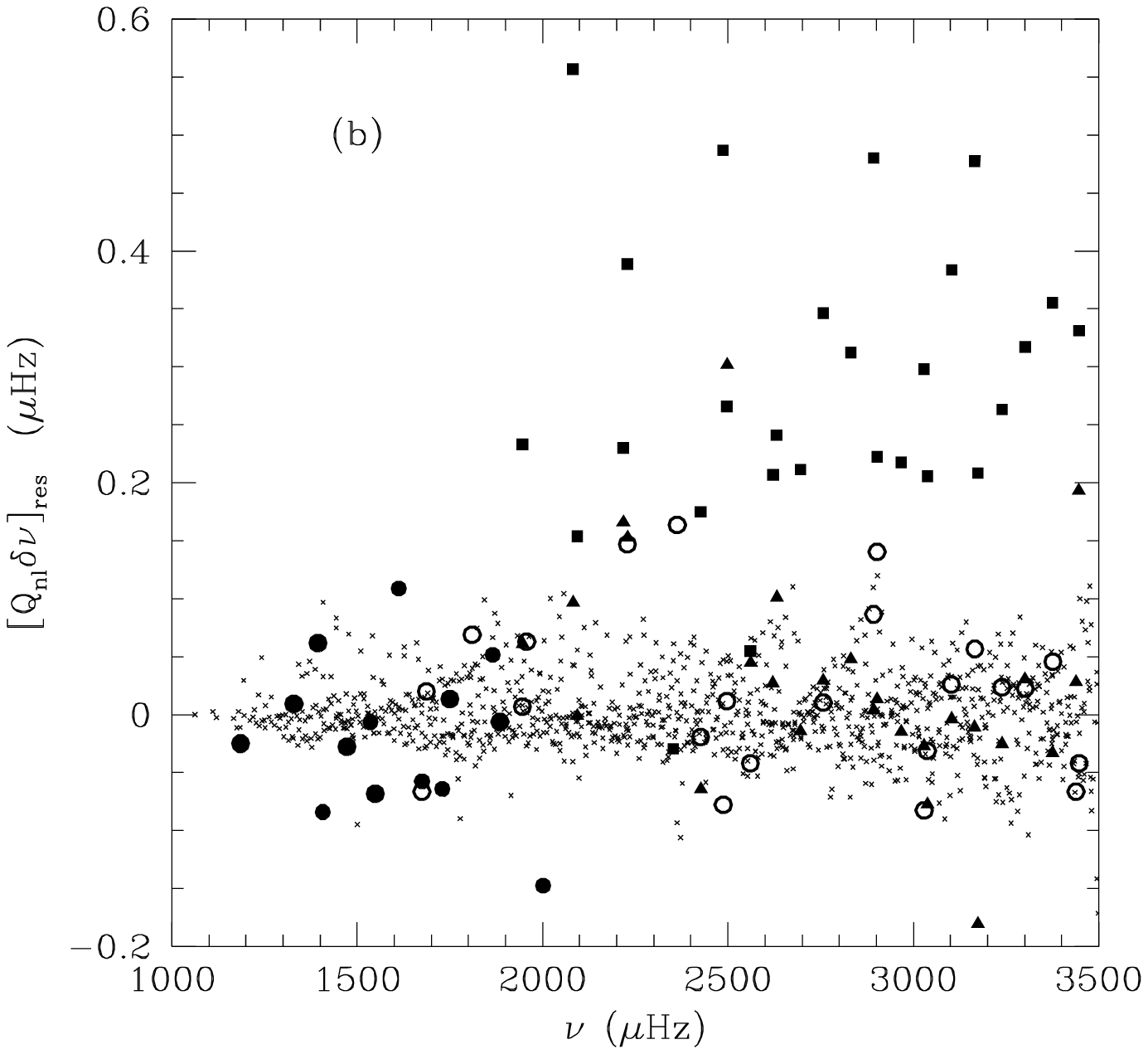}\vskip -2.0 true cm}}

\caption{\bf Figure 3. \rm (a) The frequency difference
between the different data sets and the reference model,
in the sense (Sun) -- (Model), scaled by the mode
inertia $Q_{nl}$, which is essentially the function $S$ (cf. eq.~4)
normalised by the acoustic radius of the star. (b) The residuals 
after the function $H_1(w) +  H_2(\omega)$ has 
been removed from the scaled frequency differences.
The tiny crosses are LOWL data for
degrees 3 to 10; the filled triangles are LOWL data for degrees 0--2;
the empty circles are BiSON-8 data; the filled circles are the
32-month BiSON data; and the filled squares are BiSON data obtained
at high solar activity.}

\endfigure

This effect is seen quite clearly in Fig.~3(a),
which shows differences between the
LOWL, BiSON-8 and BiSON high-activity data and
frequencies of the reference solar model.
The frequency differences have been weighted with the
corresponding mode inertia in order to compensate for the fact that,
for given a discrepancy in the model or mode physics,
deeply penetrating modes are perturbed less
than modes which have a shallow turning point 
(cf.  {\jcd} \& Berthomieu 1991).
To show the general trend of the LOWL data, we have
also plotted the frequency differences of higher-degree modes.  While
the BiSON-8, LOWL, and 32-month BiSON data follow the same trend, the
high-activity BiSON data lie above this.

Owing to the manner in which the inversion techniques are
implemented, they would be unable to isolate the two separate
frequency dependent trends resulting from the combination of the
high- and low-activity data. This would, of course, give rise to
misleading inversion results (cf. Basu et al.  1995, 1996a). 
The BiSON-8 and LOWL data show no systematic
differences, however, 
and we should therefore be able to combine these two data sets. 

Asymptotically, the frequency differences can be written as 
(e.g. Christensen-Dalsgaard, Gough \& P\'erez Hern\'an\-dez 1988)
$$
 S(w)\delta\omega/\omega=H_1(w)+H_2(\omega),\eqno\eqname\asymp
$$ 
where $w=\omega/(l+0.5)$, and $S(w)$ is a known function of the
reference model. The function $H_1(w)$ depends on the sound-speed
difference between the Sun and the reference model and $H_2(\omega)$
is determined by differences at the surface.  In Fig.~3(b) we have
plotted the residuals after the functions $H_1$ and $H_2$ -- each
obtained by fitting the LOWL data -- have been removed from the
frequency differences. Note that the residuals for the high-activity
BiSON data are much larger compared with those for the other sets,
and follow a completely different trend.

\beginfigure{4}
%\vskip 9.5 true cm
\hbox to 0 pt{\hskip -1.5cm
\vbox to 9.5 true cm{\vskip -0.5 true cm
\epsfysize=10.50 true cm\epsfbox{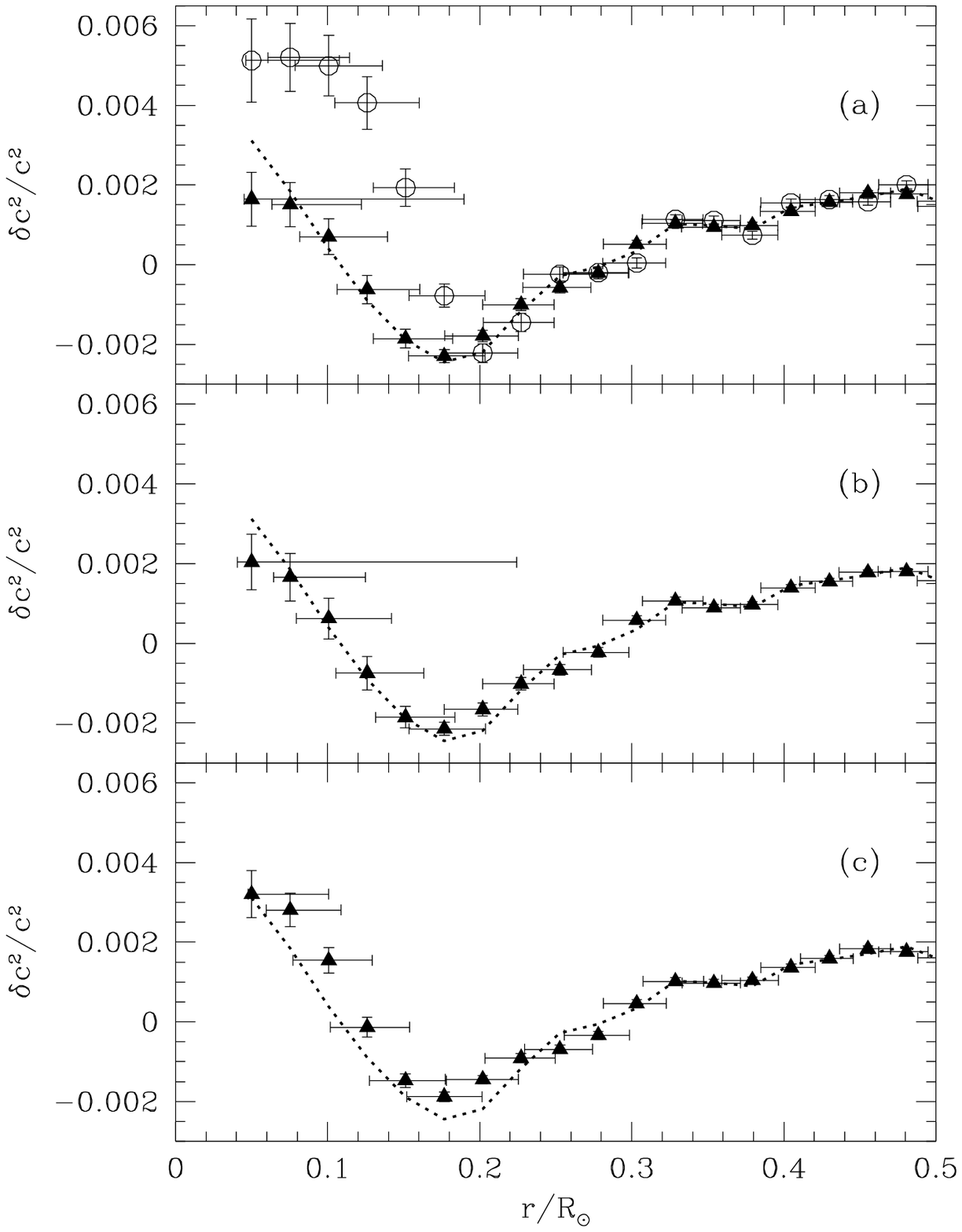}\vskip -2.0 true cm}}

\caption{\bf Figure 4. \rm Sound-speed inversion results for
different data sets, in the sense (Sun) -- (model). 
The dotted line in each plot show the results from an
inversion performed on the LOWL data only,
while the symbols are for
different combinations of LOWL and BiSON data.
The vertical $1$-$\sigma$ error bars show the propagated data errors 
and the horizontal bars extend from the first to the third
quartile point of the averaging kernels, to indicate the
resolution of the inversion.
(a) The triangles were obtained by 
using BiSON-8 data for those modes with $0 \le l \le 2$
which are common to the LOWL and BiSON-8 sets;
%the results from a set where the LOWL data
%have been substituted -- on a one-to-one basis -- with BiSON-8 data,
%i.e., only those modes that are common to both sets have been substituted; 
the circles show the results of substituting the
high-activity BiSON data in a similar fashion. 
(b) The triangles show the
results from a set where LOWL data for $0 \le l \le 2$ have been
replaced by BiSON-G data.
(c) The triangles
show the results from a set where LOWL data for $0 \le l \le 2$
have been replaced by BiSON-8 data; additional mode frequencies
from the BiSON-8 and 32-month BiSON sets -- not available in the LOWL
set -- have also been introduced.
}
\endfigure

\subsection {Consistency of inversion results}

Once the consistency of the different data sets had been established,
we proceeded to invert for the sound-speed difference between the Sun
and the reference model,
comparing results of different combinations of LOWL and BiSON data.
The results of the inversions are shown in Fig.~4. 
To provide a common reference, all panels include the results of
inverting the LOWL data only.

First, in Fig.~4(a) we
replaced the $l=0,1,\hbox{\ and\ }2$ modes of the LOWL set
with those of BiSON-8 or high-activity BiSON data. Only
those modes which were common in all sets were used -- this ensured
that the resultant resolution of the inversions would be comparable.
Since the mode sets used for the inversions were identical for
$l > 2$, any differences in the results should be confined to the core
region -- this is indeed found to be the case. As anticipated, the
results of the inversion which utilized the high-activity BiSON set
were markedly different from those based on data from the same epoch.
This
effect was first noted by Basu et al. (1996a), and is a result of
the different surface effects in the high and low-activity data.  As
can be seen in the figure, the inversion results from the LOWL set and
the LOWL plus BISON-8 set are fairly consistent -- the
contemporaneous BiSON-8 and LOWL sets can therefore be satisfactorily
combined. 
Similarly, inversion results [Fig.~4(b)]
generated by substituting the low-degree data from the 
BiSON-G set show no basic inconsistency; 
thus, although fitted by a different technique, these mode frequencies can
be used in a combined data set.

As a final step, we supplemented the input data with those low-degree
modes -- from the BiSON-8 and 32-month BiSON sets -- not present in
the LOWL set. The results are shown in Fig.~4(c). Adding the extra
modes does not change the general trend of the result,
although there are noticeable improvements in resolution and errors.

\beginfigure{5}
%\vskip 6.5 true cm
\hbox to 0 pt{\hskip -1.5cm
\vbox to 6.5 true cm{\vskip -1.75 true cm
\epsfysize=10.50 true cm\epsfbox{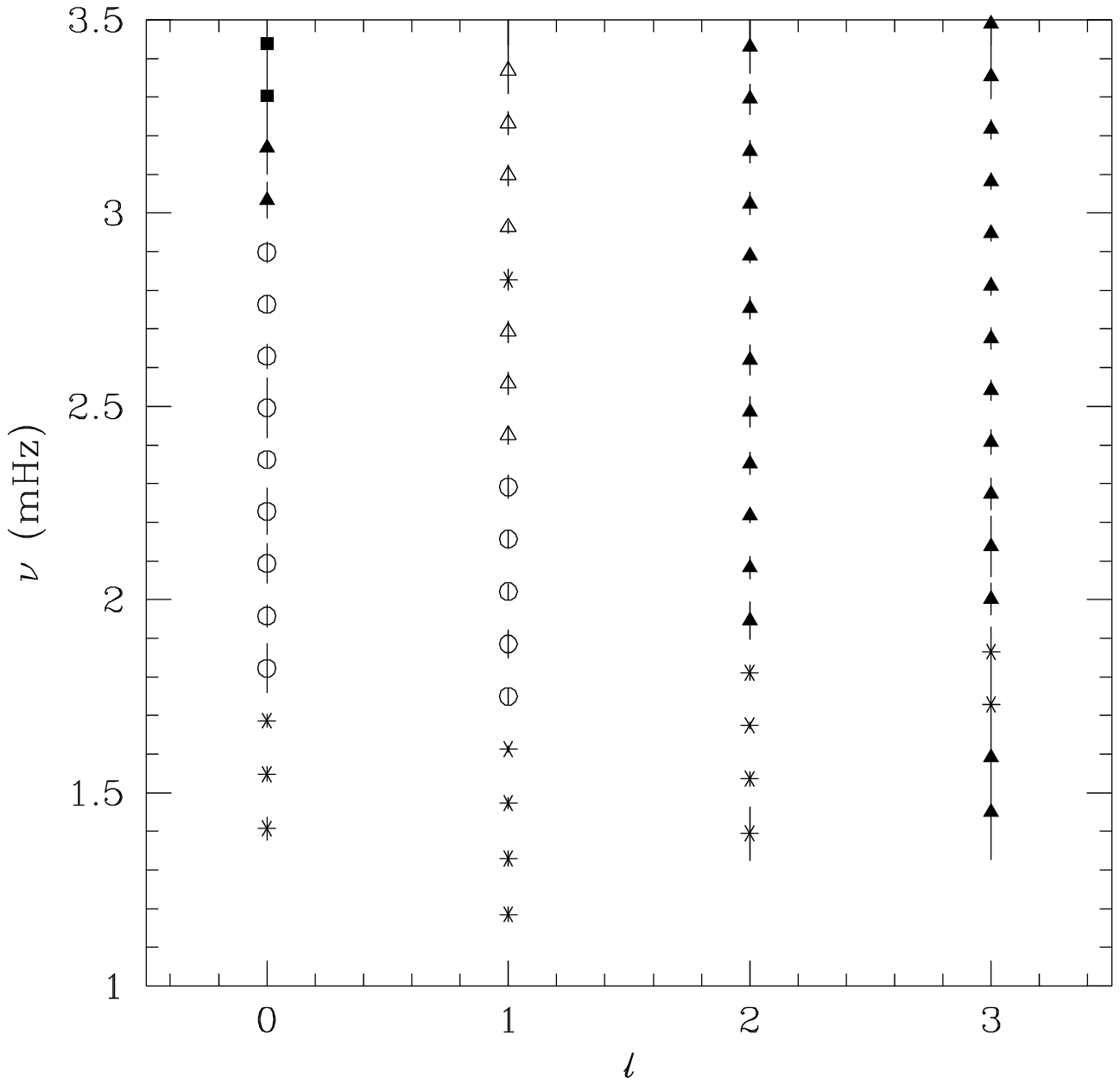}\vskip -2.0 true cm}}

\caption{\bf Figure 5. \rm The low-degree part of the $l-\nu$
diagram of the ``best'' set. The filled triangles are LOWL modes;
circles are BiSON-8 modes; asterisks are modes from the 32-month set;
filled squares the BiSON-G modes; and empty triangles averages of
LOWL and BiSON-8 modes. The error bars represent 1000$\sigma$ errors.} 

\endfigure

\subsection{The ``best'' mode set}

Having ascertained that the BiSON and LOWL data sets were compatible,
we selected a ``best'' low-degree mode set for the final inversion
results from the four contemporaneous sets (sets 1-3, 5 of Table~1).  
The mode selection
criteria were as follows.  If a mode was present in one set only it
was selected automatically.  This was true for most of the modes in
the 32-month set.
When modes were present in
both the BiSON and LOWL sets, the determination with the lower error was
selected. When the errors were essentially equal, the average frequency was
used, together with the error from just one of the sets, {\ie},
we took a conservative approach and did 
not reduce the error when averaging, on the grounds
that the errors
in the two sets were not independent (there being a correlated
contribution due to solar noise).  When faced with a choice between
the BiSON-8 and BiSON-G data, the BiSON-G data were used if the fit
to the BiSON-8 mode was obviously poor.  
For $l=4$ and above, the LOWL data were used.
The $l\le 3$ part of the $l$-$\nu$ diagram for the Best Set is shown in Fig.~5.

\beginfigure{6}
%\vskip 9.5 true cm
\moveleft 0 true cm
\vbox to 9.5 true cm{\vskip 0 true cm
\epsfysize=9.5 true cm\epsfbox{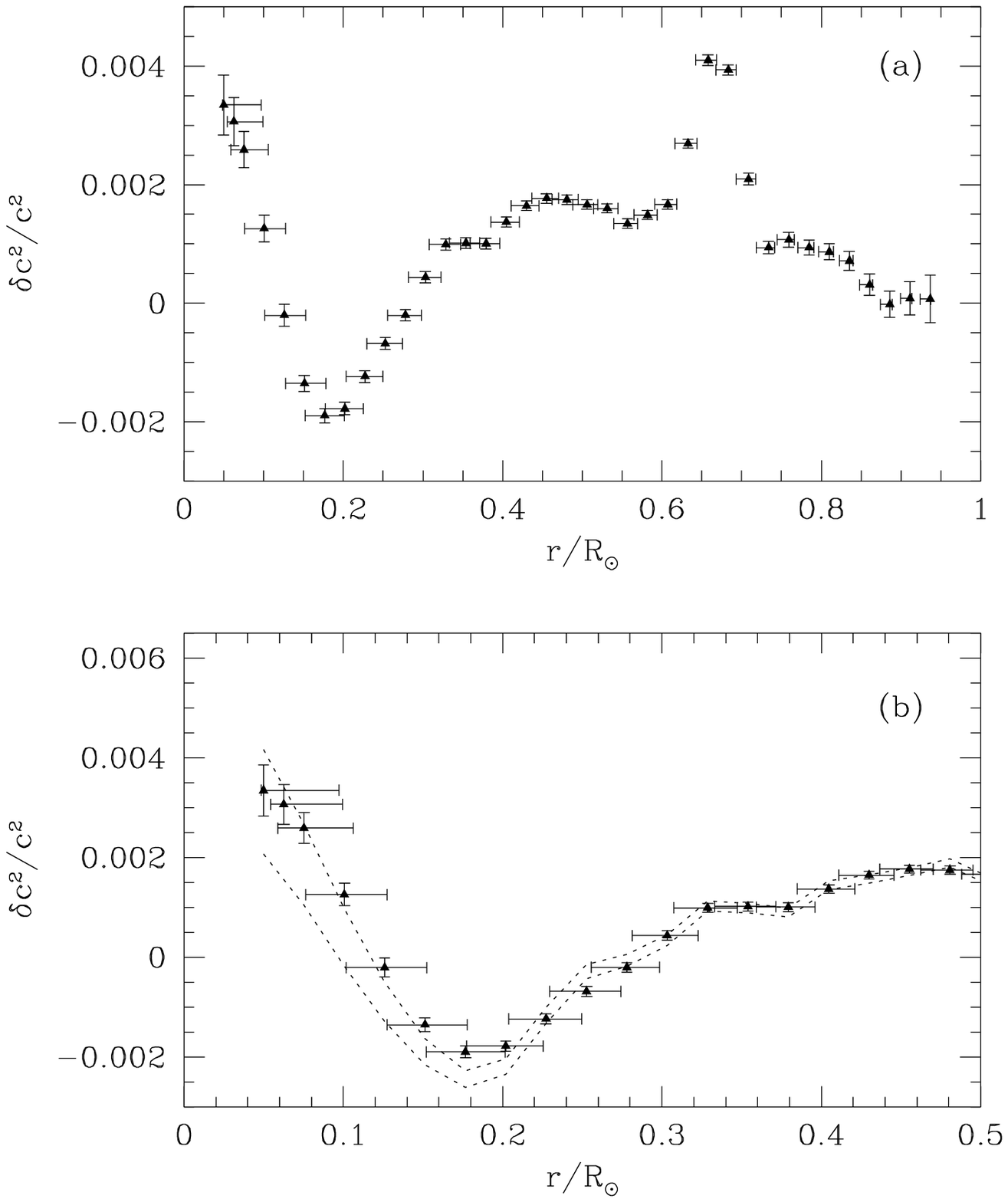}\vskip .0 true cm}
\caption{\bf Figure 6. \rm The sound-speed inversion results of the
``best'' set,
in the sense (Sun) - (Model).
The lower panel (b) concentrates on the results in the core.
For comparison, the 1$\sigma$ bounds on the result of inverting just the 
LOWL set are shown by the dotted line. The plot is on the same scale
as Fig.~4.
}
\endfigure

\beginfigure{7}
%\vskip 6.5 true cm
\hbox to 0 pt{\hskip -1.5cm
\vbox to 6.5 true cm{\vskip -1.75 true cm
\epsfysize=10.50 true cm\epsfbox{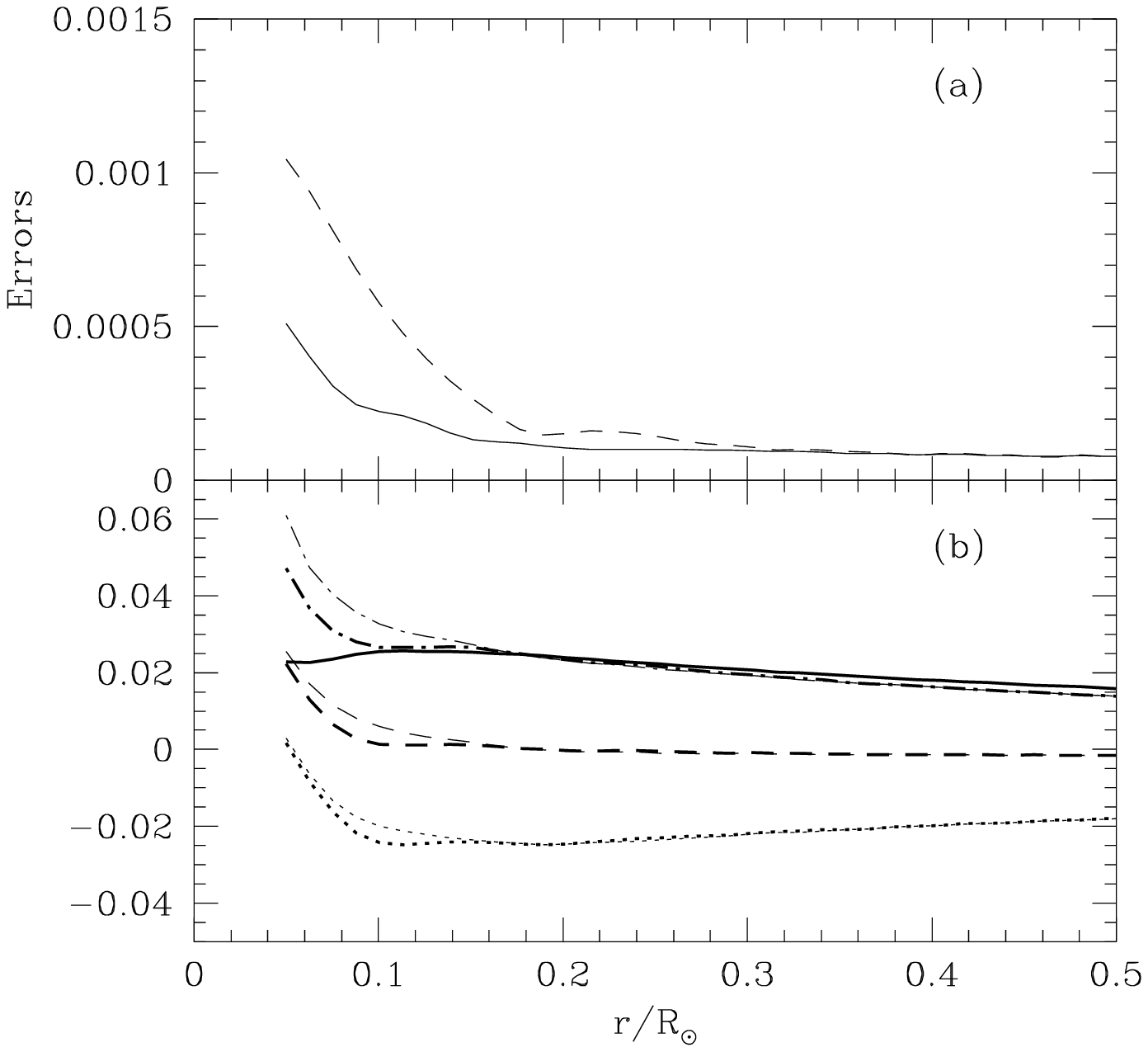}\vskip -2.0 true cm}}

\caption{\bf Figure 7. \rm A comparison of the propagated errors and
resolution of the inversion results of the LOWL and ``best'' mode
sets. The upper panel shows the $1\sigma$ errors on the solution at
different target radii. The dashed line is the error on the inversion 
results for the LOWL set, while the continuous line is that for 
the ``best'' set.
In the lower panel, the thick continuous line
is the width (in units of the solar radius $\Rsun$)
of the averaging kernels for the best set. The thick
dotted, thick dashed and thick dot-dashed line are the differences
between the first, second and third quartile points and the target
radius for the best-set inversion results, all in units of $\Rsun$. 
The thin line shows the
corresponding quantities for the inversion with the LOWL data.  The
width of the averaging kernel, defined as the distance between the
first and third quartile points.  Note that the averaging kernels
become symmetric at around 0.1$\Rsun$.}

\endfigure

\beginfigure{8}
%\vskip 8.5 true cm
\hbox to 0 pt{\hskip -2.5cm
\vbox to 8.5 true cm{\vskip -2.0 true cm
\epsfysize=12.50 true cm\epsfbox{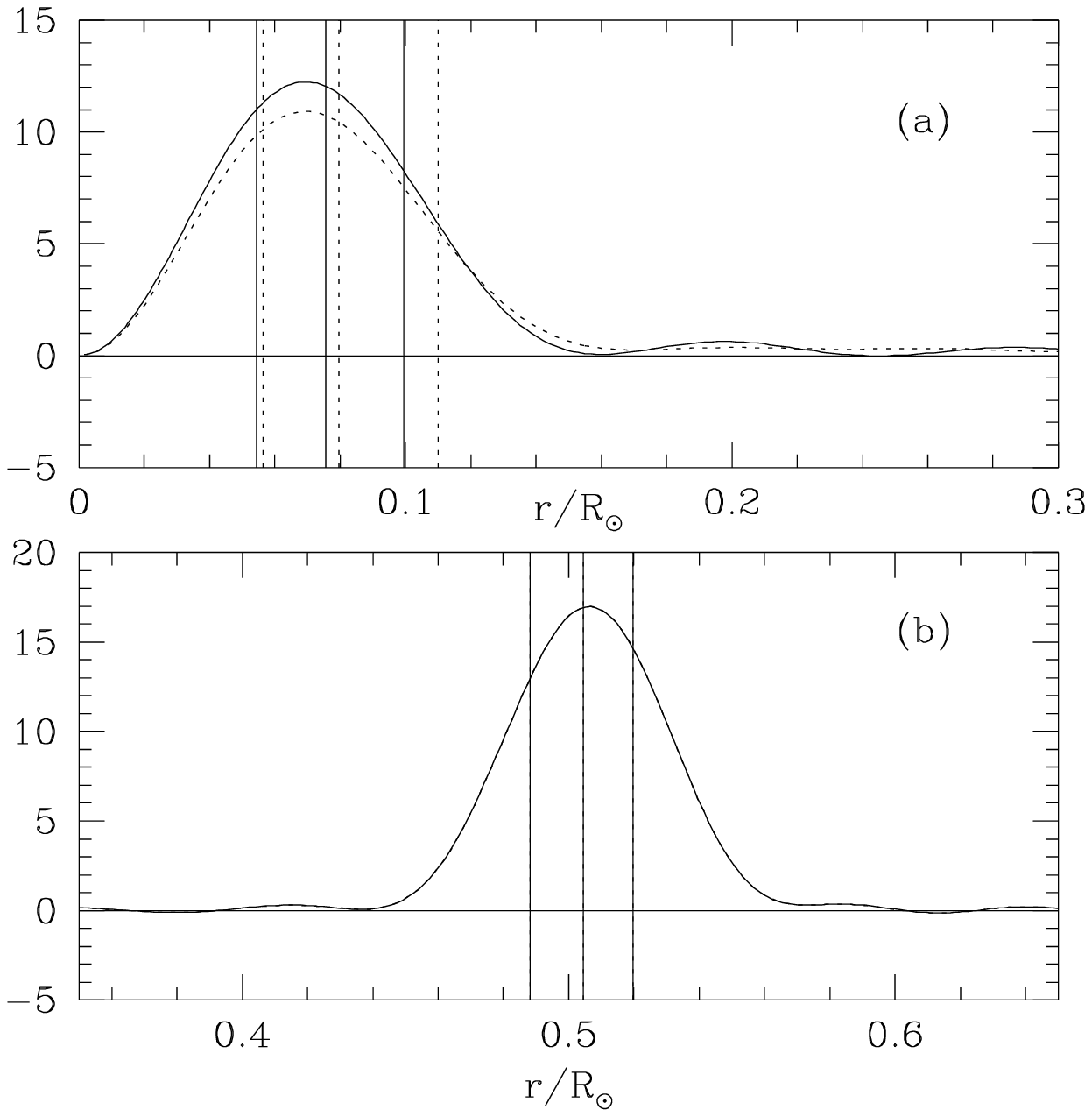}\vskip -2.0 true cm}}

\caption{\bf Figure 8. \rm The averaging kernels at target radii of
(a) 0.063 $\Rsun$ and (b) 0.51 $\Rsun$. The continuous lines
indicate the best-set inversion and the dotted lines the LOWL
inversion.  Marked on the averaging kernels are the three quartile
points. The asymmetry between the positions of the quartile points
indicated that the averaging kernels have some structure away from
the target radius.}

\endfigure

\beginfigure{9}
%\vskip 8.5 true cm
\hbox to 0 pt{\hskip -2.5cm
\vbox to 8.5 true cm{\vskip -2.0 true cm
\epsfysize=12.50 true cm\epsfbox{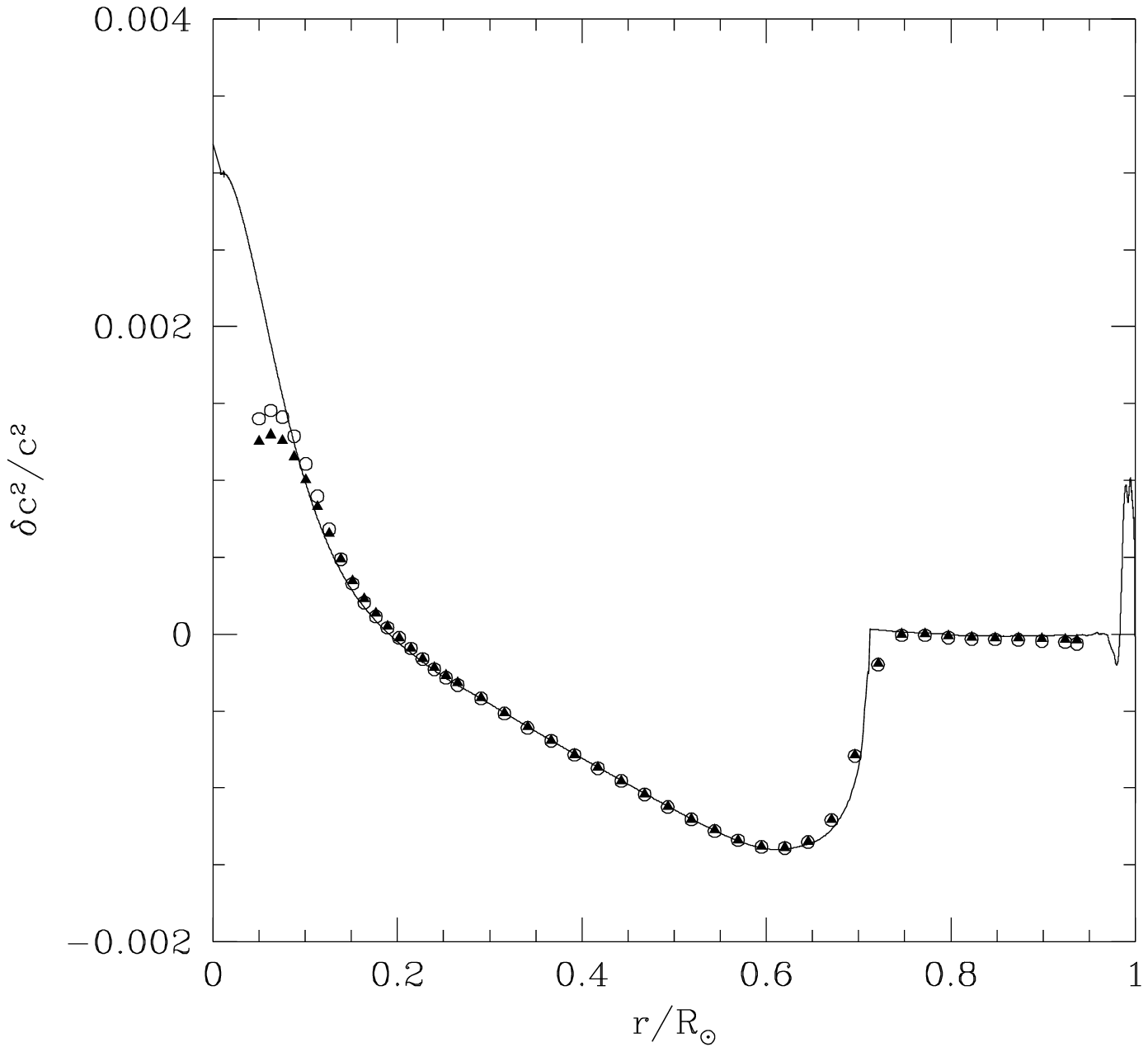}\vskip -2.0 true cm}}

\caption{\bf Figure 9. \rm The continuous line shows the 
exact sound-speed difference between the test (4.52 Gyr) and 
reference (4.6 Gyr) models, in the sense (younger) minus (older).
The symbols show the results of the inversion to determine
the difference between models. The
circles are inversion
results using the Best Set modes and the triangles those using
the LOWL modes. Note that near the core, the results using the
Best Set are closer to the exact difference. This indicates
better resolution.} 

\endfigure

The sound-speed inversion results of the Best Set are shown in
Fig.~6.  The main change in the inversion, over the LOWL-only
calculation, is that the propagated errors in the core inversion have
been reduced substantially, i.e., by more than a factor of two at
many radii.  The propagated errors in the
inversion for the LOWL-only set and Best Set are compared in Fig.~7.
Note that the influence of the low-degree modes on the errors is
restricted to radii less than 0.35 $\Rsun$.  A few averaging
kernels have been shown in Fig.~8.

As can be seen from Fig.~7, the data combination also results in a
slight improvement in resolution. This can be clearly seen by
inverting the frequency differences between the reference and the
test model -- using only the available modes in each set and
weighting by the observed errors. The results of the inversions with
the Best Set and LOWL-only set are shown in Fig.~9.  The sound-speed
difference between the two models increases towards the core. The
inverted difference is an average of the exact difference over a
finite radius -- consequently, unless the averaging is performed over
a narrow region, the inverted difference cannot match the exact
difference. Thus a higher-resolution inversion will give a better
match to the exact differences.  While the results are, as expected,
identical in the outer layers, the inversion results in the core are
not. The results of the Best Set are marginally closer to the exact
model differences than are those of the LOWL set alone.

\beginfigure{10}
%\vskip 8.5 true cm
\hbox to 0 pt{\hskip -2.5cm
\vbox to 8.5 true cm{\vskip -2.0 true cm
\epsfysize=12.50 true cm\epsfbox{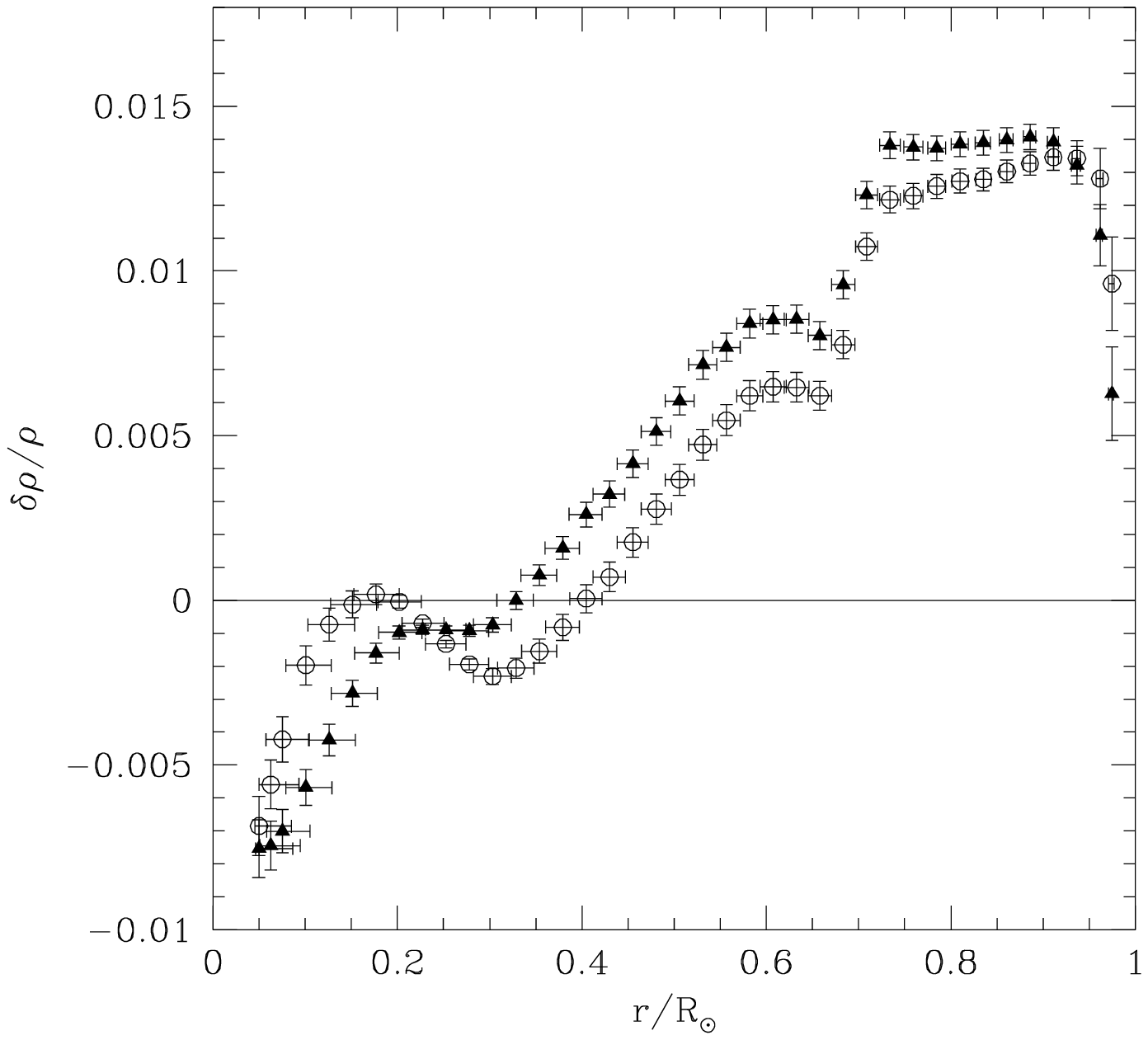}\vskip -2.0 true cm}}

\caption{\bf Figure 10. \rm The relative density difference between
the Sun and the reference model, in the sense
(Sun) -- (Model), as inferred by the inversion of the
LOWL (triangles) and the ``best'' sets (circles).}

\endfigure

We have also carried out inversions to determine the relative
density difference between the Sun and the reference model,
using the pair of variables $(\rho, Y)$.
%The combination of variables used was the
%density and the helium abundance $Y$.
The results, using both the LOWL-only set and Best Set,
are shown in Fig.~10.
Unlike the sound-speed inversion results, we find that the
density results for the two data sets differ at all radii,
even though the sets differ only in low-degree modes which predominantly 
carry information about the core.
Also, the differences in $\delta \rho/\rho$ are substantially
larger than the estimated random errors in the results.
Probably the explanation for these properties is that
our $(\rho, Y)$ inversion is sensitive
to discrepancies between the reference model's equation of state, which
was used in deriving the $(\rho,Y)$ kernels, and the true
equation of state of the solar plasma
({\cf} Basu \& Christensen-Dalsgaard 1997).
Note that in this context these discrepancies can also include the
effect on $\Gamma_1$ of differences in the abundance of heavy elements.
In additional tests with artificial data we have found 
comparable effects of using the two different mode sets
if the test model was based on the so-called MHD
equation of state (e.g. Mihalas, D\"appen \& Hummer 1988)
while the reference model used the OPAL equation of state.
It seems credible that the intrinsic error in the OPAL equation
of state, relative to the Sun, 
is of a magnitude similar to the difference between OPAL and MHD.
We also note that the mass-conservation constraint, 
$$ 
 \int \rho{\delta \rho\over\rho} r^2  \d r = 0 \; ,
 \eqno\eqname\cons
$$ 
implies a strong correlation between the results of density
inversions at all radii.
Thus even if the change in mode set has its primary
effect in the core, the inferred density difference is likely
to change at all radii, as observed. Such sensitivity to the equation 
of state can be greatly reduced by using $(\rho,\Gamma_1)$ or by explicitly 
seeking to suppress in the $(\rho,Y)$ inversion any contributions from
intrinsic differences in $\Gamma_1$ (Basu \& Christensen-Dalsgaard 1997),
although this results in larger random errors in the solution for
reasons we have already discussed in Section~1.

As in the case of the sound-speed inversion,
the errors in the density inversion are reduced and the resolution
improved by using the Best Set, compared with inversion of
the LOWL-only set.
Also, the tests on artificial data suggest that the effects
of errors in the equation of state are less severe for the
Best Set than for LOWL set, probably as a result of the
stronger constraints on the core provided by the former set.

\section{Discussion and conclusions}

We have shown that solar-oscillation data from different sources can
be successfully combined if the observations are contemporaneous.
Although a homogeneous dataset is preferable, different instruments
are best suited to observing different ranges of modes.
In particular, observations made in integrated
sunlight -- which are sensitive only to modes of the lowest degrees
-- can be combined with resolved observations to give a
reasonably large selection of modes.

Using contemporaneous data means that corrections for solar
cycle effects -- which are still very uncertain -- are not required.
The results obtained from such a combined set are consequently more
reliable.  The combination of data from the LOWL instrument and the
BiSON network shows that the inversion errors in the core can be
reduced by a up to a factor of two in the part of the Sun whose
structure is the most difficult to determine. The resolution of the
inversions is also improved to a certain extent. 

We find from Fig.~6 that the sound speed $c$ of a solar model
constructed with the most up-do-date physics available is
very close to that of the Sun. The difference in $c^2$ is at most 
0.5 per cent and is considerably less in
most regions. However, this small discrepancy is quite substantial
compared with the errors in the inversion,
as inferred by propagating the data errors.
The sound-speed difference between the Sun and the model
has three very noticeable features --- a large positive bump at
roughly $0.7\Rsun$, a dip around $0.2\Rsun$ and a rise in the
core.  The first feature is just beneath the base of the solar
convection zone, which lies at roughly $0.7 \Rsun$
({\jcd}, Gough \& Thompson 1991; Basu \& Antia 1997).
The sound speed is given by $c^2 = \Gamma_1 p /\rho \propto \Gamma_1 T/\mu$
in the excellent approximation of an ideal gas.
Thus the presence of the first feature could be explained
by helium settling being too strong in the model in this region. The
increase in the helium abundance increases the mean molecular weight
of the material at the convection-zone base, and hence decreases the
sound speed in the model. If the gravitational settling has been 
overestimated, the reason could be that in the Sun the settling is
inhibited by some
mixing in this region (Gough et al. 1996), a speculation supported by
secondary inversions for the helium profile in the Sun (Antia \&
Chitre 1997).  A mismatch in the depth
of the convection zones in the model and Sun (Basu \& Antia 1997)
could also contribute to the feature.

Probing the structure of the energy-generating core provides a
particularly important
observational test of the theory of stellar evolution.
The negative region and
the subsequent rise in the sound-speed difference implies that
the variation of the sound speed in this region of the Sun 
is flatter than
in the reference model. In this region the abundance varies
with position not only because of
settling but also due to the burning of hydrogen into helium
over the Sun's lifetime.  The inversion results could be explained if 
some mixing has transported helium outward from the inner core, leading to 
the Sun's sound speed being increased in the inner core and decreased
around $0.2R$, relative to the model
(cf. Gough et al. 1996).

The density differences between the Sun and the reference model,
illustrated in Fig.~10, are also of modest size, although
substantial compared with the estimated errors.
Overall, the density in the Sun is higher than that in
the model, except in the core -- the Sun is therefore less 
centrally condensed than the model.
The density difference shows a jump at the base of
the convection zone, corresponding to the sharp
feature in the sound-speed difference at this location.
Density differences within the convection zone
are more sensitive to model parameters than the sound speed. Such
discrepancies could be caused by differences in convection zone
depths and elemental abundances.
The results for density show a considerable, and somewhat
worrying, sensitivity to the choice of mode set.
We have argued that might plausibly be caused by errors
in the equation of state assumed in the calculation of the
reference model.
Such effects can in fact be suppressed in the inversion,
although at the expense of a substantial increase in the random
errors (cf. Basu \& Christensen-Dalsgaard 1997).
On a more positive note, the sensitivity hints at the potential
to extract even quite subtle information about the equation of state
from helioseismic inversions.

The positive sound-speed difference in the core and the smaller
central condensation 
might indicate that the Sun is younger than the assumed age of 4.6 Gyr.
Indeed, estimates based on meteoritic ages (e.g. Guenther 1989;
Wasserburg, in Bahcall \& Pinsonneault 1995)
indicate that the age is close to 4.52 Gyr, as assumed
in the test model.
Comparison of the exact difference in 
Fig.~9 with the inversion in Fig.~6 shows that using a younger
reference model would tend to reduce the difference between
the Sun and the model very near the centre, but it would
increase it elsewhere.
Similarly, Basu et al. (1996a) found that
although the density difference between the Sun and a younger model
is lower in the core, the differences elsewhere increase.
Alternatively, some degree of mixing in the core could resolve
both the problem of the larger sound speed and the lower density in
the solar core.

\section*{Acknowledgments}

This work was supported in part by the Danish National Research
Foundation through its establishment of the Theoretical Astrophysics
Center and by the US National Science foundation through base funding of
HAO/NCAR. We would like to thank all those who are -- or have been --
associated with the BiSON global network and all our hosts at
the network sites.  BiSON is funded by the UK
Particle Physics and Astronomy Research Council.

%\notea[Steve, Jesper -- please add your respective acknows.]

\section*{References}

%\notea[still needs checking]

\bye

%% file: mn.tex
% MN.TEX (Computer Modern version)
%
% plain TeX single / double column macros for the
% Monthly Notices of Royal Astronomical Society
%
% v1.6  (mn.tex)  --- released 18th September 1995 (A. Woollatt)
% v1.5      "     --- released 25th August 1994 (M. Reed)
% v1.4      "     --- released 22nd February 1994
% v1.3  (mnd.tex) --- released 28th November 1992
% v1.26     "     --- released  1st August 1992
% v1.25     "     --- released 25th February 1992
%
% Copyright Cambridge University Press
%
% > Incorporating special symbol code from laa.sty v1.1 (25th Feb 1991)
%   used with the permission of Springer Verlag.
% > Incorporating parts of mssymb.tex (8th July 1987).
% > Incorporating NewFont.sty v ALPHA patchlevel 8 (16th August 1994).
% > Add footlines, add footnotes in double column (18th September
%   1995).

\catcode `\@=11 % @ signs are letters

\def\@version{1.6}
\def\@verdate{18th September 1995}

% Fonts: Computer Modern / Monotype Times (CUP only)
%
% Font family sizes available:
%   8pt, 9pt, 10pt, 11pt, 14pt and 17pt.
%
% Faces available:
%   \rm, math italic, symbol, \it, \bf, \sl, \tt, \sc, \sf, \cal, \em,
%   \mit and \oldstyle.

% define the typeface in use

\newif\ifprod@font

\ifx\@typeface\undefined
  \def\@typeface{Comp. Modern}\prod@fontfalse
\else
  \prod@fonttrue % We want Times
\fi

\def\newfam{\alloc@8\fam\chardef\sixt@@n} % made not outer

\ifprod@font
\font\fiverm=mtr10 at 5pt
\font\fivebf=mtbx10 at 5pt
\font\fiveit=mtti10 at 5pt
\font\fivesl=mtsl10 at 5pt
\font\fivett=cmtt8 at 5pt     \hyphenchar\fivett=-1
\font\fivecsc=mtcsc10 at 5pt
\font\fivesf=mtss10 at 5pt
\font\fivei=mtmi10 at 5pt      \skewchar\fivei='177
\font\fivesy=mtsy10 at 5pt     \skewchar\fivesy='60

\font\sixrm=mtr10 at 6pt
\font\sixbf=mtbx10 at 6pt
\font\sixit=mtti10 at 6pt
\font\sixsl=mtsl10 at 6pt
\font\sixtt=cmtt8 at 6pt      \hyphenchar\sixtt=-1
\font\sixcsc=mtcsc10 at 6pt
\font\sixsf=mtss10 at 6pt
\font\sixi=mtmi10 at 6pt       \skewchar\sixi='177
\font\sixsy=mtsy10 at 6pt      \skewchar\sixsy='60

\font\sevenrm=mtr10 at 7pt
\font\sevenbf=mtbx10 at 7pt
\font\sevenit=mtti10 at 7pt
\font\sevensl=mtsl10 at 7pt
\font\seventt=cmtt8 at 7pt     \hyphenchar\seventt=-1
\font\sevencsc=mtcsc10 at 7pt
\font\sevensf=mtss10 at 7pt
\font\seveni=mtmi10 at 7pt      \skewchar\seveni='177
\font\sevensy=mtsy10 at 7pt     \skewchar\sevensy='60

\font\eightrm=mtr10 at 8pt
\font\eightbf=mtbx10 at 8pt
\font\eightit=mtti10 at 8pt
\font\eighti=mtmi10 at 8pt      \skewchar\eighti='177
\font\eightsy=mtsy10 at 8pt     \skewchar\eightsy='60
\font\eightsl=mtsl10 at 8pt
\font\eighttt=cmtt8             \hyphenchar\eighttt=-1
\font\eightcsc=mtcsc10 at 8pt
\font\eightsf=mtss10 at 8pt

\font\ninerm=mtr10 at 9pt
\font\ninebf=mtbx10 at 9pt
\font\nineit=mtti10 at 9pt
\font\ninei=mtmi10 at 9pt      \skewchar\ninei='177
\font\ninesy=mtsy10 at 9pt     \skewchar\ninesy='60
\font\ninesl=mtsl10 at 9pt
\font\ninett=cmtt9             \hyphenchar\ninett=-1
\font\ninecsc=mtcsc10 at 9pt
\font\ninesf=mtss10 at 9pt

\font\tenrm=mtr10
\font\tenbf=mtbx10
\font\tenit=mtti10
\font\teni=mtmi10		\skewchar\teni='177
\font\tensy=mtsy10		\skewchar\tensy='60
\font\tenex=cmex10
\font\tensl=mtsl10
\font\tentt=cmtt10		\hyphenchar\tentt=-1
\font\tencsc=mtcsc10
\font\tensf=mtss10

\font\elevenrm=mtr10 at 11pt
\font\elevenbf=mtbx10 at 11pt
\font\elevenit=mtti10 at 11pt
\font\eleveni=mtmi10 at 11pt      \skewchar\eleveni='177
\font\elevensy=mtsy10 at 11pt     \skewchar\elevensy='60
\font\elevensl=mtsl10 at 11pt
\font\eleventt=cmtt10 at 11pt     \hyphenchar\eleventt=-1
\font\elevencsc=mtcsc10 at 11pt
\font\elevensf=mtss10 at 11pt

\font\twelverm=mtr10 at 12pt
\font\twelvebf=mtbx10 at 12pt
\font\twelveit=mtti10 at 12pt
\font\twelvesl=mtsl10 at 12pt
\font\twelvett=cmtt12             \hyphenchar\twelvett=-1
\font\twelvecsc=mtcsc10 at 12pt
\font\twelvesf=mtss10 at 12pt
\font\twelvei=mtmi10 at 12pt      \skewchar\twelvei='177
\font\twelvesy=mtsy10 at 12pt     \skewchar\twelvesy='60

\font\fourteenrm=mtr10 at 14pt
\font\fourteenbf=mtbx10 at 14pt
\font\fourteenit=mtti10 at 14pt
\font\fourteeni=mtmi10 at 14pt      \skewchar\fourteeni='177
\font\fourteensy=mtsy10 at 14pt     \skewchar\fourteensy='60
\font\fourteensl=mtsl10 at 14pt
\font\fourteentt=cmtt12 at 14pt     \hyphenchar\fourteentt=-1
\font\fourteencsc=mtcsc10 at 14pt
\font\fourteensf=mtss10 at 14pt

\font\seventeenrm=mtr10 at 17pt
\font\seventeenbf=mtbx10 at 17pt
\font\seventeenit=mtti10 at 17pt
\font\seventeeni=mtmi10 at 17pt      \skewchar\seventeeni='177
\font\seventeensy=mtsy10 at 17pt     \skewchar\seventeensy='60
\font\seventeensl=mtsl10 at 17pt
\font\seventeentt=cmtt12 at 17pt     \hyphenchar\seventeentt=-1
\font\seventeencsc=mtcsc10 at 17pt
\font\seventeensf=mtss10 at 17pt
\else
\font\fiverm=cmr5
\font\fivei=cmmi5             \skewchar\fivei='177
\font\fivesy=cmsy5            \skewchar\fivesy='60
\font\fivebf=cmbx5

\font\sixrm=cmr6
\font\sixi=cmmi6             \skewchar\sixi='177
\font\sixsy=cmsy6            \skewchar\sixsy='60
\font\sixbf=cmbx6

\font\sevenrm=cmr7
\font\sevenit=cmti7
\font\seveni=cmmi7             \skewchar\seveni='177
\font\sevensy=cmsy7            \skewchar\sevensy='60
\font\sevenbf=cmbx7

\font\eightrm=cmr8
\font\eightbf=cmbx8
\font\eightit=cmti8
\font\eighti=cmmi8			\skewchar\eighti='177
\font\eightsy=cmsy8			\skewchar\eightsy='60
\font\eightsl=cmsl8
\font\eighttt=cmtt8			\hyphenchar\eighttt=-1
\font\eightcsc=cmcsc10 at 8pt
\font\eightsf=cmss8

\font\ninerm=cmr9
\font\ninebf=cmbx9
\font\nineit=cmti9
\font\ninei=cmmi9			\skewchar\ninei='177
\font\ninesy=cmsy9			\skewchar\ninesy='60
\font\ninesl=cmsl9
\font\ninett=cmtt9			\hyphenchar\ninett=-1
\font\ninecsc=cmcsc10 at 9pt
\font\ninesf=cmss9

\font\tenrm=cmr10
\font\tenbf=cmbx10
\font\tenit=cmti10
\font\teni=cmmi10		\skewchar\teni='177
\font\tensy=cmsy10		\skewchar\tensy='60
\font\tenex=cmex10
\font\tensl=cmsl10
\font\tentt=cmtt10		\hyphenchar\tentt=-1
\font\tencsc=cmcsc10
\font\tensf=cmss10

\font\elevenrm=cmr10 scaled \magstephalf
\font\elevenbf=cmbx10 scaled \magstephalf
\font\elevenit=cmti10 scaled \magstephalf
\font\eleveni=cmmi10 scaled \magstephalf	\skewchar\eleveni='177
\font\elevensy=cmsy10 scaled \magstephalf	\skewchar\elevensy='60
\font\elevensl=cmsl10 scaled \magstephalf
\font\eleventt=cmtt10 scaled \magstephalf	\hyphenchar\eleventt=-1
\font\elevencsc=cmcsc10 scaled \magstephalf
\font\elevensf=cmss10 scaled \magstephalf

\font\twelverm=cmr10 scaled \magstep1
\font\twelvebf=cmbx10 scaled \magstep1
\font\twelvei=cmmi10 scaled \magstep1      \skewchar\twelvei='177
\font\twelvesy=cmsy10 scaled \magstep1     \skewchar\twelvesy='60

\font\fourteenrm=cmr10 scaled \magstep2
\font\fourteenbf=cmbx10 scaled \magstep2
\font\fourteenit=cmti10 scaled \magstep2
\font\fourteeni=cmmi10 scaled \magstep2		\skewchar\fourteeni='177
\font\fourteensy=cmsy10 scaled \magstep2	\skewchar\fourteensy='60
\font\fourteensl=cmsl10 scaled \magstep2
\font\fourteentt=cmtt10 scaled \magstep2	\hyphenchar\fourteentt=-1
\font\fourteencsc=cmcsc10 scaled \magstep2
\font\fourteensf=cmss10 scaled \magstep2

\font\seventeenrm=cmr10 scaled \magstep3
\font\seventeenbf=cmbx10 scaled \magstep3
\font\seventeenit=cmti10 scaled \magstep3
\font\seventeeni=cmmi10 scaled \magstep3	\skewchar\seventeeni='177
\font\seventeensy=cmsy10 scaled \magstep3	\skewchar\seventeensy='60
\font\seventeensl=cmsl10 scaled \magstep3
\font\seventeentt=cmtt10 scaled \magstep3	\hyphenchar\seventeentt=-1
\font\seventeencsc=cmcsc10 scaled \magstep3
\font\seventeensf=cmss10 scaled \magstep3
\fi

\def\hexnumber#1{\ifcase#1 0\or1\or2\or3\or4\or5\or6\or7\or8\or9\or
  A\or B\or C\or D\or E\or F\fi}

\def\makestrut{%
  \setbox\strutbox=\hbox{%
    \vrule height.7\baselineskip depth.3\baselineskip width \z@}%
}

\def\baselinestretch{1}
\newskip\tmp@bls

\def\b@ls#1{% set baseline using \baselinestretch as a scale factor
  \tmp@bls=#1\relax
  \baselineskip=#1\relax\makestrut
  \normalbaselineskip=\baselinestretch\tmp@bls
  \normalbaselines
}

\def\nostb@ls#1{% set baseline skip ignoring \baselinestretch
  \normalbaselineskip=#1\relax
  \normalbaselines
  \makestrut
}

% families \itfam, \slfam, \bffam, \ttfam defined in PLAIN.
%
% \itfam is \fam4
% \slfam is \fam5
% \bffam is \fam6
% \ttfam is \fam7

\newfam\scfam  % \fam8
\newfam\sffam  % \fam9

\def\mit{\fam\@ne}
\def\cal{\fam\tw@}
\def\em{\ifdim\fontdimen1\font>\z@ \rm\else\it\fi}

\textfont3=\tenex
\scriptfont3=\tenex
\scriptscriptfont3=\tenex

\setbox0=\hbox{\tenex B} \p@renwd=\wd0 % width of the big left (

\def\eightpoint{% 8^6^5 on 10pt
  \def\rm{\fam0\eightrm}%
  \textfont0=\eightrm \scriptfont0=\sixrm \scriptscriptfont0=\fiverm%
  \textfont1=\eighti  \scriptfont1=\sixi  \scriptscriptfont1=\fivei%
  \textfont2=\eightsy \scriptfont2=\sixsy \scriptscriptfont2=\fivesy%
  \textfont\itfam=\eightit\def\it{\fam\itfam\eightit}%
  \ifprod@font
    \scriptfont\itfam=\sixit
      \scriptscriptfont\itfam=\fiveit
  \else
    \scriptfont\itfam=\eightit
      \scriptscriptfont\itfam=\eightit
  \fi
  \textfont\bffam=\eightbf%
    \scriptfont\bffam=\sixbf%
      \scriptscriptfont\bffam=\fivebf%
  \def\bf{\fam\bffam\eightbf}%
  \textfont\slfam=\eightsl\def\sl{\fam\slfam\eightsl}%
  \ifprod@font
    \scriptfont\slfam=\sixsl
      \scriptscriptfont\slfam=\fivesl
  \else
    \scriptfont\slfam=\eightsl
      \scriptscriptfont\slfam=\eightsl
  \fi
  \textfont\ttfam=\eighttt\def\tt{\fam\ttfam\eighttt}%
  \ifprod@font
    \scriptfont\ttfam=\sixtt
      \scriptscriptfont\ttfam=\fivett
  \else
    \scriptfont\ttfam=\eighttt
      \scriptscriptfont\ttfam=\eighttt
  \fi
  \textfont\scfam=\eightcsc\def\sc{\fam\scfam\eightcsc}%
  \ifprod@font
    \scriptfont\scfam=\sixcsc
      \scriptscriptfont\scfam=\fivecsc
  \else
    \scriptfont\scfam=\eightcsc
      \scriptscriptfont\scfam=\eightcsc
  \fi
  \textfont\sffam=\eightsf\def\sf{\fam\sffam\eightsf}%
  \ifprod@font
    \scriptfont\sffam=\sixsf
      \scriptscriptfont\sffam=\fivesf
  \else
    \scriptfont\sffam=\eightsf
      \scriptscriptfont\sffam=\eightsf
  \fi
  \def\oldstyle{\fam\@ne\eighti}%
  \b@ls{10pt}\rm\@viiipt%
}
\def\@viiipt{}

\def\ninepoint{% 9^6^5 on 11pt (two col) / 12 (single col)
  \def\rm{\fam0\ninerm}%
  \textfont0=\ninerm \scriptfont0=\sixrm \scriptscriptfont0=\fiverm%
  \textfont1=\ninei  \scriptfont1=\sixi  \scriptscriptfont1=\fivei%
  \textfont2=\ninesy \scriptfont2=\sixsy \scriptscriptfont2=\fivesy%
  \textfont\itfam=\nineit\def\it{\fam\itfam\nineit}%
  \ifprod@font
    \scriptfont\itfam=\sixit
      \scriptscriptfont\itfam=\fiveit
  \else
    \scriptfont\itfam=\nineit
      \scriptscriptfont\itfam=\nineit
  \fi
  \textfont\bffam=\ninebf%
    \scriptfont\bffam=\sixbf%
      \scriptscriptfont\bffam=\fivebf%
  \def\bf{\fam\bffam\ninebf}%
  \textfont\slfam=\ninesl\def\sl{\fam\slfam\ninesl}%
  \ifprod@font
    \scriptfont\slfam=\sixsl
      \scriptscriptfont\slfam=\fivesl
  \else
    \scriptfont\slfam=\ninesl
      \scriptscriptfont\slfam=\ninesl
  \fi
  \textfont\ttfam=\ninett\def\tt{\fam\ttfam\ninett}%
  \ifprod@font
    \scriptfont\ttfam=\sixtt
      \scriptscriptfont\ttfam=\fivett
  \else
    \scriptfont\ttfam=\ninett
      \scriptscriptfont\ttfam=\ninett
  \fi
  \textfont\scfam=\ninecsc\def\sc{\fam\scfam\ninecsc}%
  \ifprod@font
    \scriptfont\scfam=\sixcsc
      \scriptscriptfont\scfam=\fivecsc
  \else
    \scriptfont\scfam=\ninecsc
      \scriptscriptfont\scfam=\ninecsc
  \fi
  \textfont\sffam=\ninesf\def\sf{\fam\sffam\ninesf}%
  \ifprod@font
    \scriptfont\sffam=\sixsf
      \scriptscriptfont\sffam=\fivesf
  \else
    \scriptfont\sffam=\ninesf
      \scriptscriptfont\sffam=\ninesf
  \fi
  \def\oldstyle{\fam\@ne\ninei}%
  \b@ls{\TextLeading plus \Feathering}\rm\@ixpt%
}
\def\@ixpt{}

\def\tenpoint{% 10^7^5 on 11pt
  \def\rm{\fam0\tenrm}%
  \textfont0=\tenrm \scriptfont0=\sevenrm \scriptscriptfont0=\fiverm%
  \textfont1=\teni  \scriptfont1=\seveni  \scriptscriptfont1=\fivei%
  \textfont2=\tensy \scriptfont2=\sevensy \scriptscriptfont2=\fivesy%
  \textfont\itfam=\tenit\def\it{\fam\itfam\tenit}%
  \ifprod@font
    \scriptfont\itfam=\sevenit
      \scriptscriptfont\itfam=\fiveit
  \else
    \scriptfont\itfam=\tenit
      \scriptscriptfont\itfam=\tenit
  \fi
  \textfont\bffam=\tenbf%
    \scriptfont\bffam=\sevenbf%
      \scriptscriptfont\bffam=\fivebf%
  \def\bf{\fam\bffam\tenbf}%
  \textfont\slfam=\tensl\def\sl{\fam\slfam\tensl}%
  \ifprod@font
    \scriptfont\slfam=\sevensl
      \scriptscriptfont\slfam=\fivesl
  \else
    \scriptfont\slfam=\tensl
      \scriptscriptfont\slfam=\tensl
  \fi
  \textfont\ttfam=\tentt\def\tt{\fam\ttfam\tentt}%
  \ifprod@font
    \scriptfont\ttfam=\seventt
      \scriptscriptfont\ttfam=\fivett
  \else
    \scriptfont\ttfam=\tentt
      \scriptscriptfont\ttfam=\tentt
  \fi
  \textfont\scfam=\tencsc\def\sc{\fam\scfam\tencsc}%
  \ifprod@font
    \scriptfont\scfam=\sevencsc
      \scriptscriptfont\scfam=\fivecsc
  \else
    \scriptfont\scfam=\tencsc
      \scriptscriptfont\scfam=\tencsc
  \fi
  \textfont\sffam=\tensf\def\sf{\fam\sffam\tensf}%
  \ifprod@font
    \scriptfont\sffam=\sevensf
      \scriptscriptfont\sffam=\fivesf
  \else
    \scriptfont\sffam=\tensf
      \scriptscriptfont\sffam=\tensf
  \fi
  \def\oldstyle{\fam\@ne\teni}%
  \b@ls{11pt}\rm\@xpt%
}
\def\@xpt{}

\def\elevenpoint{% 11^8^6 on 13pt
  \def\rm{\fam0\elevenrm}%
  \textfont0=\elevenrm \scriptfont0=\eightrm \scriptscriptfont0=\sixrm%
  \textfont1=\eleveni  \scriptfont1=\eighti  \scriptscriptfont1=\sixi%
  \textfont2=\elevensy \scriptfont2=\eightsy \scriptscriptfont2=\sixsy%
  \textfont\itfam=\elevenit\def\it{\fam\itfam\elevenit}%
  \ifprod@font
    \scriptfont\itfam=\eightit
      \scriptscriptfont\itfam=\sixit
  \else
    \scriptfont\itfam=\elevenit
      \scriptscriptfont\itfam=\elevenit
  \fi
  \textfont\bffam=\elevenbf%
    \scriptfont\bffam=\eightbf%
      \scriptscriptfont\bffam=\sixbf%
  \def\bf{\fam\bffam\elevenbf}%
  \textfont\slfam=\elevensl\def\sl{\fam\slfam\elevensl}%
  \ifprod@font
    \scriptfont\slfam=\eightsl
      \scriptscriptfont\slfam=\sixsl
  \else
    \scriptfont\slfam=\elevensl
      \scriptscriptfont\slfam=\elevensl
  \fi
  \textfont\ttfam=\eleventt\def\tt{\fam\ttfam\eleventt}%
  \ifprod@font
    \scriptfont\ttfam=\eighttt
      \scriptscriptfont\ttfam=\sixtt
  \else
    \scriptfont\ttfam=\eleventt
      \scriptscriptfont\ttfam=\eleventt
  \fi
  \textfont\scfam=\elevencsc\def\sc{\fam\scfam\elevencsc}%
  \ifprod@font
    \scriptfont\scfam=\eightcsc
      \scriptscriptfont\scfam=\sixcsc
  \else
    \scriptfont\scfam=\elevencsc
      \scriptscriptfont\scfam=\elevencsc
  \fi
  \textfont\sffam=\elevensf\def\sf{\fam\sffam\elevensf}%
  \ifprod@font
    \scriptfont\sffam=\eightsf
      \scriptscriptfont\sffam=\sixsf
  \else
    \scriptfont\sffam=\elevensf
      \scriptscriptfont\sffam=\elevensf
  \fi
  \def\oldstyle{\fam\@ne\eleveni}%
  \b@ls{13pt}\rm\@xipt%
}
\def\@xipt{}

\def\fourteenpoint{% 14^10^7 on 17pt
  \def\rm{\fam0\fourteenrm}%
  \textfont0\fourteenrm  \scriptfont0\tenrm  \scriptscriptfont0\sevenrm%
  \textfont1\fourteeni   \scriptfont1\teni   \scriptscriptfont1\seveni%
  \textfont2\fourteensy  \scriptfont2\tensy  \scriptscriptfont2\sevensy%
  \textfont\itfam=\fourteenit\def\it{\fam\itfam\fourteenit}%
  \ifprod@font
    \scriptfont\itfam=\tenit
      \scriptscriptfont\itfam=\sevenit
  \else
    \scriptfont\itfam=\fourteenit
      \scriptscriptfont\itfam=\fourteenit
  \fi
  \textfont\bffam=\fourteenbf%
    \scriptfont\bffam=\tenbf%
      \scriptscriptfont\bffam=\sevenbf%
  \def\bf{\fam\bffam\fourteenbf}%
  \textfont\slfam=\fourteensl\def\sl{\fam\slfam\fourteensl}%
  \ifprod@font
    \scriptfont\slfam=\tensl
      \scriptscriptfont\slfam=\sevensl
  \else
    \scriptfont\slfam=\fourteensl
      \scriptscriptfont\slfam=\fourteensl
  \fi
  \textfont\ttfam=\fourteentt\def\tt{\fam\ttfam\fourteentt}%
  \ifprod@font
    \scriptfont\ttfam=\tentt
      \scriptscriptfont\ttfam=\seventt
  \else
    \scriptfont\ttfam=\fourteentt
      \scriptscriptfont\ttfam=\fourteentt
  \fi
  \textfont\scfam=\fourteencsc\def\sc{\fam\scfam\fourteencsc}%
  \ifprod@font
    \scriptfont\scfam=\tencsc
      \scriptscriptfont\scfam=\sevencsc
  \else
    \scriptfont\scfam=\fourteencsc
      \scriptscriptfont\scfam=\fourteencsc
  \fi
  \textfont\sffam=\fourteensf\def\sf{\fam\sffam\fourteensf}%
  \ifprod@font
    \scriptfont\sffam=\tensf
      \scriptscriptfont\sffam=\sevensf
  \else
    \scriptfont\sffam=\fourteensf
      \scriptscriptfont\sffam=\fourteensf
  \fi
  \def\oldstyle{\fam\@ne\fourteeni}%
  \b@ls{17pt}\rm\@xivpt%
}
\def\@xivpt{}

\def\seventeenpoint{% 17^12^10 on 20pt
  \def\rm{\fam0\seventeenrm}%
  \textfont0\seventeenrm  \scriptfont0\twelverm  \scriptscriptfont0\tenrm%
  \textfont1\seventeeni   \scriptfont1\twelvei   \scriptscriptfont1\teni%
  \textfont2\seventeensy  \scriptfont2\twelvesy  \scriptscriptfont2\tensy%
  \textfont\itfam=\seventeenit\def\it{\fam\itfam\seventeenit}%
  \ifprod@font
    \scriptfont\itfam=\twelveit
      \scriptscriptfont\itfam=\tenit
  \else
    \scriptfont\itfam=\seventeenit
      \scriptscriptfont\itfam=\seventeenit
  \fi
  \textfont\bffam=\seventeenbf%
    \scriptfont\bffam=\twelvebf%
      \scriptscriptfont\bffam=\tenbf%
  \def\bf{\fam\bffam\seventeenbf}%
  \textfont\slfam=\seventeensl\def\sl{\fam\slfam\seventeensl}%
  \ifprod@font
    \scriptfont\slfam=\twelvesl
      \scriptscriptfont\slfam=\tensl
  \else
    \scriptfont\slfam=\seventeensl
      \scriptscriptfont\slfam=\seventeensl
  \fi
  \textfont\ttfam=\seventeentt\def\tt{\fam\ttfam\seventeentt}%
  \ifprod@font
    \scriptfont\ttfam=\twelvett
      \scriptscriptfont\ttfam=\tentt
  \else
    \scriptfont\ttfam=\seventeentt
      \scriptscriptfont\ttfam=\seventeentt
  \fi
  \textfont\scfam=\seventeencsc\def\sc{\fam\scfam\seventeencsc}%
  \ifprod@font
    \scriptfont\scfam=\twelvecsc
      \scriptscriptfont\scfam=\tencsc
  \else
    \scriptfont\scfam=\seventeencsc
      \scriptscriptfont\scfam=\seventeencsc
  \fi
  \textfont\sffam=\seventeensf\def\sf{\fam\sffam\seventeensf}%
  \ifprod@font
    \scriptfont\sffam=\twelvesf
      \scriptscriptfont\sffam=\tensf
  \else
    \scriptfont\sffam=\seventeensf
      \scriptscriptfont\sffam=\seventeensf
  \fi
  \def\oldstyle{\fam\@ne\seventeeni}%
  \b@ls{20pt}\rm\@xviipt%
}
\def\@xviipt{}

\lineskip=1pt      \normallineskip=\lineskip
\lineskiplimit=\z@ \normallineskiplimit=\lineskiplimit

% BOLD MATH SYMBOLS

% Make \, work in non-math mode
\def\,{\relax\ifmmode \mskip\thinmuskip\else \thinspace\fi}
\let\protect=\relax

\long\def\@ifundefined#1#2#3{\expandafter\ifx\csname
  #1\endcsname\relax#2\else#3\fi}

%%%%%%%%%%%%%%%%%%%%%%%%%%%%%%%%%%%%%%%%%

% NewFont.sty: ALPHA VERSION patchlevel 8, 16th August 1994, M. Reed

% \addtom@thgroup{math font loading info}
% Adds to internal \math@groups definition, which is executed at the end
% of each size changing command. It is called by \NewSymbolFont.

\newtoks\math@groups \math@groups={}
\def\addtom@thgroup#1#2{#1\expandafter{\the#1#2}} %  \mac={new\the\mac}

% Make TeX change the values of \s@ze, \ss@ze, \sss@ze when \@npt is
% executed. This makes it possible for math characters to be loaded
% at the correct size automatically when the size is changed.

% \addtosizeh@ok{x}{10}{7}{5}

\def\addtosizeh@ok#1#2#3#4{%
  \expandafter\def\csname @#1pt\endcsname{%
    \def\s@ze{#2}\def\ss@ze{#3}\def\sss@ze{#4}\the\math@groups%
  }%
}

% \resetsizehook allows the size parameters to be reset after \addtosizeh@ok
% has been called (it re-defines \@npt).
% e.g JFM which requires \xpt to have 10.5pt instead of 10pt.
% Note: \resetsizehook must be used in the preamble BEFORE any
% \New... commands.

% e.g. \resetsizehook{x}{10.5}{7}{5}

\let\resetsizehook=\addtosizeh@ok

% Standard LaTeX sizes

\ifprod@font
%  \addtosizeh@ok{v}    {5} {5}  {5}
%  \addtosizeh@ok{vi}   {6} {6}  {6}
%  \addtosizeh@ok{vii}  {7} {6}  {5}
  \addtosizeh@ok{viii} {8} {6}  {5}
  \addtosizeh@ok{ix}   {9} {6}  {5}
  \addtosizeh@ok{x}    {10}{7}  {5}
  \addtosizeh@ok{xi}   {11}{8}  {6}
%  \addtosizeh@ok{xii}  {12}{8}  {6}
  \addtosizeh@ok{xiv}  {14}{10} {7}
  \addtosizeh@ok{xvii} {17}{12}{10}
%  \addtosizeh@ok{xx}   {20}{14}{12}
%  \addtosizeh@ok{xxv}  {25}{20}{17}
\else
%  \addtosizeh@ok{v}    {5}     {5}     {5}
%  \addtosizeh@ok{vi}   {6}     {6}     {6}
%  \addtosizeh@ok{vii}  {7}     {6}     {5}
  \addtosizeh@ok{viii} {8}     {6}     {5}
  \addtosizeh@ok{ix}   {9}     {6}     {5}
  \addtosizeh@ok{x}    {10}    {7}     {5}
  \addtosizeh@ok{xi}   {10.95} {8}     {6}
%  \addtosizeh@ok{xii}  {12}    {8}     {6}
  \addtosizeh@ok{xiv}  {14.4}  {10}    {7}
  \addtosizeh@ok{xvii} {17.28} {12}    {10}
%  \addtosizeh@ok{xx}   {20.74} {14.4}  {12}
%  \addtosizeh@ok{xxv}  {24.88} {20.74} {17.28}
\fi

\def\get@font#1#2#3{%
  \edef\fonts@ze{\romannumeral#3}%         10 -> x
  \edef\fontn@me{\fonts@ze#1}%             AMSa -> xAMSa
  \@ifundefined{\fontn@me}%
    {%%\typeout{defining \fontn@me}%
     \global\expandafter\font\csname \fontn@me\endcsname=#2 at #3pt}%
    {}%
}

\def\ass@tfont#1#2{%
  \xdef\fam@name{\csname #1\endcsname}%
  \xdef\font@name{\csname #2\endcsname}%
  \let\textfont@name\font@name
  \textfont\fam@name\textfont@name
}

\def\ass@sfont#1#2{%
  \xdef\fam@name{\csname #1\endcsname}%
  \xdef\font@name{\csname #2\endcsname}%
  \let\textfont@name\font@name
  \scriptfont\fam@name\textfont@name
}

\def\ass@ssfont#1#2{%
  \xdef\fam@name{\csname #1\endcsname}%
  \xdef\font@name{\csname #2\endcsname}%
  \let\textfont@name\font@name
  \scriptscriptfont\fam@name\textfont@name
}

%                fam name  base font  (allocates a \newfam)
% \NewSymbolFont {AMSa}    {mtxm10}

\def\NewSymbolFont#1#2{%
  \expandafter\ifx\csname sym#1fam\endcsname\relax % if not defined
    \expandafter\newfam\csname sym#1fam\endcsname
    \expandafter\edef\csname sym#1fam\endcsname{\the\allocationnumber}%
    \addtom@thgroup\math@groups{%
      \get@font{#1}{#2}{\s@ze}%
      \ass@tfont{sym#1fam}{\fontn@me}%
      \get@font{#1}{#2}{\ss@ze}%
      \ass@sfont{sym#1fam}{\fontn@me}%
      \get@font{#1}{#2}{\sss@ze}%
      \ass@ssfont{sym#1fam}{\fontn@me}%
    }%
  \else
    \errmessage{Family `#1' already defined}%
  \fi
}

%                symbol         type fam    pos (hex)
% \NewMathSymbol {\blacksquare} {0}  {AMSa} {04}

\def\NewMathSymbol#1#2#3#4{%
  \edef\f@mly{\expandafter\hexnumber{\csname sym#3fam\endcsname}}%
  \mathchardef#1="#2\f@mly#4\relax
}

%                  macro name  type  fam1   pos  fam2   pos
% \NewMathDelimiter{\ulcorner} {4}   {AMSa} {70} {AMSb} {70}

\newif\ifd@f

\def\NewMathDelimiter#1#2#3#4#5#6{%
  \d@ftrue
  \expandafter\ifx\csname sym#3fam\endcsname\relax
    \d@ffalse \errmessage{Family `#3' is not defined}%
  \fi
  \expandafter\ifx\csname sym#5fam\endcsname\relax
    \d@ffalse \errmessage{Family `#5' is not defined}%
  \fi
  \ifd@f
    \edef\f@mly{\expandafter\hexnumber{\csname sym#3fam\endcsname}}%
    \edef\f@mlytw@{\expandafter\hexnumber{\csname sym#5fam\endcsname}}%
    \xdef#1{\delimiter"#2\f@mly #4\f@mlytw@ #6\relax}%
  \fi
}

%                  macro name  base font  skewchar setting e.g '60 (octal)
% \NewMathAlphabet {mathbssi}  {mtmisb10} {}

\def\setboxz@h{\setbox\z@\hbox}
\def\wdz@{\wd\z@}
\def\boxz@{\box\z@}
\def\setbox@ne{\setbox\@ne}
\def\wd@ne{\wd\@ne}

\def\math@atom#1#2{%
   \binrel@{#1}\binrel@@{#2}}
\def\binrel@#1{\setboxz@h{\thinmuskip0mu
  \medmuskip\m@ne mu\thickmuskip\@ne mu$#1\m@th$}%
 \setbox@ne\hbox{\thinmuskip0mu\medmuskip\m@ne mu\thickmuskip
  \@ne mu${}#1{}\m@th$}%
 \setbox\tw@\hbox{\hskip\wd@ne\hskip-\wdz@}}
\def\binrel@@#1{\ifdim\wd2<\z@\mathbin{#1}\else\ifdim\wd\tw@>\z@
 \mathrel{#1}\else{#1}\fi\fi}

\def\m@thit{1}

\def\set@skchar#1{\global\expandafter\skewchar
  \csname\fontn@me\endcsname=#1\relax}

\def\NewMathAlphabet#1#2#3{%
  \def\tst{#3}%
  \ifx\tst\empty\else % if a \skewchar setting is present
    \expandafter\gdef\csname #1@sc\endcsname{}%  \def\cmd@sc{..}
  \fi
  \expandafter\def\csname #1\endcsname{%  \def\cmd{\protect\@cmd}
    \protect\csname @#1\endcsname}%
  \expandafter\def\csname @#1\endcsname##1{%  \def\@cmd{..}
    {%
    \begingroup
      \get@font{#1}{#2}{\s@ze}%
      \@ifundefined{#1@sc}{}{\set@skchar{#3}}%
      \ass@tfont{m@thit}{\fontn@me}%
      \get@font{#1}{#2}{\ss@ze}%
      \@ifundefined{#1@sc}{}{\set@skchar{#3}}%
      \ass@sfont{m@thit}{\fontn@me}%
      \get@font{#1}{#2}{\sss@ze}%
      \@ifundefined{#1@sc}{}{\set@skchar{#3}}%
      \ass@ssfont{m@thit}{\fontn@me}%
      \math@atom{##1}{%
      \mathchoice%
        {\hbox{$\m@th\displaystyle##1$}}%
        {\hbox{$\m@th\textstyle##1$}}%
        {\hbox{$\m@th\scriptstyle##1$}}%
        {\hbox{$\m@th\scriptscriptstyle##1$}}}%
    \endgroup
    }%
  }%
}

%                  macro name  base font  hyphenchar setting e.g -1 (off)
% \NewTextAlphabet {textbfit}  {mtbxti10} {}

% save a family if \NewTextAlphabet is not used.
\newif\iffirstta  \firsttatrue

\def\set@hchar#1{\global\expandafter\hyphenchar
  \csname\fontn@me\endcsname=#1\relax}

\def\NewTextAlphabet#1#2#3{%
  \iffirstta
    \global\firsttafalse
    \newfam\scratchfam
    \edef\scrt@fam{\the\allocationnumber}%
  \fi
  \def\tst{#3}%
  \ifx\tst\empty\else % if a \hyphenchar setting is required
    \expandafter\gdef\csname #1@hc\endcsname{}%  \def\cmd@sc{..}
  \fi
  \expandafter\def\csname #1\endcsname{%  \def\cmd{\protect\t@cmd}
    \protect\csname t@#1\endcsname}%
  \long\expandafter\def\csname t@#1\endcsname##1{%  \def\t@cmd{..}
    \ifmmode
      \typeout{Warning: do not use \expandafter\string\csname #1\endcsname
        \space in math mode}\fi%
    {%
      \get@font{#1}{#2}{\s@ze}\let\t@xtfnt=\fontn@me\relax
      \@ifundefined{#1@hc}{}{\set@hchar{#3}}%
      \ass@tfont{scrt@fam}{\fontn@me}%
      \get@font{#1}{#2}{\ss@ze}%
      \@ifundefined{#1@hc}{}{\set@hchar{#3}}%
      \ass@sfont{scrt@fam}{\fontn@me}%
      \get@font{#1}{#2}{\sss@ze}%
      \@ifundefined{#1@hc}{}{\set@hchar{#3}}%
      \ass@ssfont{scrt@fam}{\fontn@me}%
      \fam\scratchfam\csname\t@xtfnt\endcsname
    ##1%
    }%
  }%
  \expandafter\def\csname #1shape%  \def\cmdshape{\protect\@cmdshape}
    \endcsname{\protect\csname @#1shape\endcsname}%
  \expandafter\def\csname @#1shape\endcsname{%  \def\@cmdshape
    \ifmmode
      \typeout{Warning: do not use \expandafter\string\csname
        #1shape\endcsname \space in math mode}\fi
      \get@font{#1}{#2}{\s@ze}\let\t@xtfnt=\fontn@me\relax
      \@ifundefined{#1@hc}{}{\set@hchar{#3}}%
      \ass@tfont{scrt@fam}{\fontn@me}%
      \get@font{#1}{#2}{\ss@ze}%
      \@ifundefined{#1@hc}{}{\set@hchar{#3}}%
      \ass@sfont{scrt@fam}{\fontn@me}%
      \get@font{#1}{#2}{\sss@ze}%
      \@ifundefined{#1@hc}{}{\set@hchar{#3}}%
      \ass@ssfont{scrt@fam}{\fontn@me}%
      \fam\scratchfam\csname\t@xtfnt\endcsname
  }%
}

% \bmath{math text}

\ifprod@font
  \def\math@itfnt{mtmib10}
  \def\math@syfnt{mtbsy10}
\else
  \def\math@itfnt{cmmib10}
  \def\math@syfnt{cmbsy10}
\fi

\def\m@thsy{2}

\def\bmath{\protect\@bmath}
\def\@bmath#1{%
  {%
  \begingroup
    \get@font{mthit}{\math@itfnt}{\s@ze}\set@skchar{'177}%
    \ass@tfont{m@thit}{\fontn@me}%
    \get@font{mthit}{\math@itfnt}{\ss@ze}\set@skchar{'177}%
    \ass@sfont{m@thit}{\fontn@me}%
    \get@font{mthit}{\math@itfnt}{\sss@ze}\set@skchar{'177}%
    \ass@ssfont{m@thit}{\fontn@me}%
    \get@font{mthsy}{\math@syfnt}{\s@ze}\set@skchar{'60}%
    \ass@tfont{m@thsy}{\fontn@me}%
    \get@font{mthsy}{\math@syfnt}{\ss@ze}\set@skchar{'60}%
    \ass@sfont{m@thsy}{\fontn@me}%
    \get@font{mthsy}{\math@syfnt}{\sss@ze}\set@skchar{'60}%
    \ass@ssfont{m@thsy}{\fontn@me}%
    \math@atom{#1}{%
    \mathchoice%
      {\hbox{$\m@th\displaystyle#1$}}%
      {\hbox{$\m@th\textstyle#1$}}%
      {\hbox{$\m@th\scriptstyle#1$}}%
      {\hbox{$\m@th\scriptscriptstyle#1$}}}%
  \endgroup
  }%
}

%%%%%%%%%%%%%%%%%%%%%%%%%%%%%%%%%%%%%%%%%

% Astronomy and Astrophysics symbol macros

\def\diameter{{\ifmmode\mathchoice
{\ooalign{\hfil\hbox{$\displaystyle/$}\hfil\crcr
{\hbox{$\displaystyle\mathchar"20D$}}}}
{\ooalign{\hfil\hbox{$\textstyle/$}\hfil\crcr
{\hbox{$\textstyle\mathchar"20D$}}}}
{\ooalign{\hfil\hbox{$\scriptstyle/$}\hfil\crcr
{\hbox{$\scriptstyle\mathchar"20D$}}}}
{\ooalign{\hfil\hbox{$\scriptscriptstyle/$}\hfil\crcr
{\hbox{$\scriptscriptstyle\mathchar"20D$}}}}
\else{\ooalign{\hfil/\hfil\crcr\mathhexbox20D}}%
\fi}}

\def\sq{\ifmmode\squareforqed\else{\unskip\nobreak\hfil
\penalty50\hskip1em\null\nobreak\hfil\squareforqed
\parfillskip=0pt\finalhyphendemerits=0\endgraf}\fi}
\def\squareforqed{\hbox{\rlap{$\sqcap$}$\sqcup$}}

% Simulated Blackboard Bold symbols

\def\bbbc{{\mathchoice {\setbox0=\hbox{$\displaystyle\rm C$}\hbox{\hbox
to0pt{\kern0.4\wd0\vrule height0.9\ht0\hss}\box0}}
{\setbox0=\hbox{$\textstyle\rm C$}\hbox{\hbox
to0pt{\kern0.4\wd0\vrule height0.9\ht0\hss}\box0}}
{\setbox0=\hbox{$\scriptstyle\rm C$}\hbox{\hbox
to0pt{\kern0.4\wd0\vrule height0.9\ht0\hss}\box0}}
{\setbox0=\hbox{$\scriptscriptstyle\rm C$}\hbox{\hbox
to0pt{\kern0.4\wd0\vrule height0.9\ht0\hss}\box0}}}}
\def\bbbq{{\mathchoice {\setbox0=\hbox{$\displaystyle\rm
Q$}\hbox{\raise
0.15\ht0\hbox to0pt{\kern0.4\wd0\vrule height0.8\ht0\hss}\box0}}
{\setbox0=\hbox{$\textstyle\rm Q$}\hbox{\raise
0.15\ht0\hbox to0pt{\kern0.4\wd0\vrule height0.8\ht0\hss}\box0}}
{\setbox0=\hbox{$\scriptstyle\rm Q$}\hbox{\raise
0.15\ht0\hbox to0pt{\kern0.4\wd0\vrule height0.7\ht0\hss}\box0}}
{\setbox0=\hbox{$\scriptscriptstyle\rm Q$}\hbox{\raise
0.15\ht0\hbox to0pt{\kern0.4\wd0\vrule height0.7\ht0\hss}\box0}}}}
\def\bbbt{{\mathchoice {\setbox0=\hbox{$\displaystyle\rm
T$}\hbox{\hbox to0pt{\kern0.3\wd0\vrule height0.9\ht0\hss}\box0}}
{\setbox0=\hbox{$\textstyle\rm T$}\hbox{\hbox
to0pt{\kern0.3\wd0\vrule height0.9\ht0\hss}\box0}}
{\setbox0=\hbox{$\scriptstyle\rm T$}\hbox{\hbox
to0pt{\kern0.3\wd0\vrule height0.9\ht0\hss}\box0}}
{\setbox0=\hbox{$\scriptscriptstyle\rm T$}\hbox{\hbox
to0pt{\kern0.3\wd0\vrule height0.9\ht0\hss}\box0}}}}
\def\bbbs{{\mathchoice
{\setbox0=\hbox{$\displaystyle     \rm S$}\hbox{\raise0.5\ht0\hbox
to0pt{\kern0.35\wd0\vrule height0.45\ht0\hss}\hbox
to0pt{\kern0.55\wd0\vrule height0.5\ht0\hss}\box0}}
{\setbox0=\hbox{$\textstyle        \rm S$}\hbox{\raise0.5\ht0\hbox
to0pt{\kern0.35\wd0\vrule height0.45\ht0\hss}\hbox
to0pt{\kern0.55\wd0\vrule height0.5\ht0\hss}\box0}}
{\setbox0=\hbox{$\scriptstyle      \rm S$}\hbox{\raise0.5\ht0\hbox
to0pt{\kern0.35\wd0\vrule height0.45\ht0\hss}\raise0.05\ht0\hbox
to0pt{\kern0.5\wd0\vrule height0.45\ht0\hss}\box0}}
{\setbox0=\hbox{$\scriptscriptstyle\rm S$}\hbox{\raise0.5\ht0\hbox
to0pt{\kern0.4\wd0\vrule height0.45\ht0\hss}\raise0.05\ht0\hbox
to0pt{\kern0.55\wd0\vrule height0.45\ht0\hss}\box0}}}}
\def\bbbz{{\mathchoice {\hbox{$\sf\textstyle Z\kern-0.4em Z$}}
{\hbox{$\sf\textstyle Z\kern-0.4em Z$}}
{\hbox{$\sf\scriptstyle Z\kern-0.3em Z$}}
{\hbox{$\sf\scriptscriptstyle Z\kern-0.2em Z$}}}}

% NUMBER THE DESIGN ELEMENTS

\def\Nulle{0} % null element
\def\Afe{1}   % author affiliation
\def\Hae{2}   % heading A
\def\Hbe{3}   % heading B
\def\Hce{4}   % heading C
\def\Hde{5}   % heading D

% TEMPORARY REGISTERS

\newcount\LastMac       \LastMac=\Nulle

\newskip\half      \half=5.5pt plus 1.5pt minus 2.25pt
\newskip\one       \one=11pt plus 3pt minus 5.5pt
\newskip\onehalf   \onehalf=16.5pt plus 5.5pt minus 8.25pt
\newskip\two       \two=22pt plus 5.5pt minus 11pt

\def\Half{\addvspace{\half}}
\def\One{\addvspace{\one}}
\def\OneHalf{\addvspace{\onehalf}}
\def\Two{\addvspace{\two}}

\def\Raggedright{% set lines unjustified
  \rightskip=\z@ plus \hsize\relax
}

\def\Fullout{% set lines justified
  \rightskip=\z@\relax
}

\def\Hang#1#2{% set hanging indentation
  \hangindent=#1%
  \hangafter=#2\relax
}

% Pagestyles

\newif\ifsp@page
\def\pagestyle#1{\csname ps@#1\endcsname}
\def\thispagestyle#1{\global\sp@pagetrue\gdef\sp@type{#1}}

\def\ps@titlepage{%
  \def\@oddhead{\eightpoint\noindent \the\CatchLine
    \ifprod@font\else\qquad Printed\ \today\qquad
      (MN plain \TeX\ macros\ v\@version)\fi \hfil}%
  \let\@evenhead=\@oddhead
  \def\@oddfoot{\eightpoint\copyright\ \@pubyear\ RAS\hfil}%
  \def\@evenfoot{\hfil\eightpoint\noindent\copyright\ \@pubyear\ RAS}%
}

\def\ps@headings{%
  \def\@oddhead{\elevenpoint\it\noindent
    \hfill\the\RightHeader\hskip1.5em\rm\folio}%
  \def\@evenhead{\elevenpoint\noindent
    \folio\hskip1.5em\it\the\LeftHeader\hfill}%
  \def\@oddfoot{\eightpoint\noindent\copyright\ \@pubyear\ RAS,
    MNRAS {\bf \@volume}, \@pagerange\hfil}%
  \def\@evenfoot{\hfil\eightpoint\copyright\ \@pubyear\ RAS,
    MNRAS {\bf \@volume}, \@pagerange}%
}

\def\ps@plate{%
  \def\@oddhead{\eightpoint\noindent\plt@cap\hfil}%
  \def\@evenhead{\eightpoint\noindent\plt@cap\hfil}%
  \def\@oddfoot{\eightpoint\noindent\copyright\ \@pubyear\ RAS,
    MNRAS {\bf \@volume}, \@pagerange\hfil}%
  \def\@evenfoot{\hfil\eightpoint\copyright\ \@pubyear\ RAS,
    MNRAS {\bf \@volume}, \@pagerange}%
}

% DESIGN ELEMENT DEFINITIONS

% Article opening

\def\title#1{% article title
  \bgroup
    \vbox to 8pt{\vss}%
    \seventeenpoint
    \Raggedright
    \noindent \strut{\bf #1}\par
  \egroup
}

\def\author#1{% article author(s)
  \bgroup
    \ifnum\LastMac=\Afe \OneHalf\else \vskip 21pt\fi
    \fourteenpoint
    \Raggedright
    \noindent \strut #1\par
    \vskip 3pt%
  \egroup
}

\def\affiliation#1{% author(s) affiliation
  \bgroup
    \vskip -4pt%
    \eightpoint
    \Raggedright
    \noindent \strut {\it #1}\par
  \egroup
  \LastMac=\Afe\relax
}

\def\acceptedline#1{% acceptance date
  \bgroup
    \Two
    \eightpoint
    \Raggedright
    \noindent \strut #1\par
  \egroup
}

\long\def\abstract#1{%
  \bgroup
    \vskip 20pt%
    \leftskip 11pc\rightskip\z@
    \noindent{\ninebf ABSTRACT}\par
    \tenpoint
    \Fullout
    \noindent #1\par
  \egroup
}

\long\def\keywords#1{% keywords
  \bgroup
    \Half
    \leftskip 11pc\rightskip\z@
    \tenpoint
    \Fullout
    \noindent\hbox{\bf Key words:}\ #1\par
  \egroup
}

% The \maketitle macro ensures that the two spanning material appears
% at the top of the first page, and that it has two lines of space
% underneath it. If you forget this in you input, no output will be produced.
% The \BeginOpening (alias \begintopmatter) macro should be called at the
% very start of the input file, so that it is in force when the document
% starts. This ensures that when \maketitle is called that the group is
% closed, and the material gets printed.

\def\maketitle{%
  \EndOpening
  \ifsinglecol \else \MakePage\fi
}

% Page offset

% Counter setup

\def\@nameuse#1{\csname #1\endcsname}
\def\arabic#1{\@arabic{\@nameuse{#1}}}
\def\alph#1{\@alph{\@nameuse{#1}}}
\def\Alph#1{\@Alph{\@nameuse{#1}}}
\def\@arabic#1{\number #1}
\def\@Alph#1{\ifcase#1\or A\or B\or C\or D\else\@Ialph{#1}\fi}
\def\@Ialph#1{\ifcase#1\or \or \or \or \or E\or F\or G\or H\or I\or J\or
   K\or L\or M\or N\or O\or P\or Q\or R\or S\or T\or U\or V\or W\or X\or
   Y\or Z\else\errmessage{Counter out of range}\fi}
\def\@alph#1{\ifcase#1\or a\or b\or c\or d\else\@ialph{#1}\fi}
\def\@ialph#1{\ifcase#1\or \or \or \or \or e\or f\or g\or h\or i\or j\or
   k\or l\or m\or n\or o\or p\or q\or r\or s\or t\or u\or v\or w\or x\or y\or
   z\else\errmessage{Counter out of range}\fi}

% Equation auto-numbering

\newcount\Eqnno
\newcount\SubEqnno

\def\theeq{\arabic{Eqnno}}
\def\thesubeq{\alph{SubEqnno}}

\def\stepeq{\relax
  \global\SubEqnno \z@
  \global\advance\Eqnno \@ne\relax
  {\rm (\theeq)}%
}

\def\startsubeq{\relax
  \global\SubEqnno \z@
  \global\advance\Eqnno \@ne\relax
  \stepsubeq
}

\def\stepsubeq{\relax
  \global\advance\SubEqnno \@ne\relax
  {\rm (\theeq\thesubeq)}%
}

% Headings

\newcount\Sec        %  heading auto number counters
\newcount\SecSec
\newcount\SecSecSec

\def\thesection{\arabic{Sec}}
\def\thesubsection{\thesection.\arabic{SecSec}}
\def\thesubsubsection{\thesubsection.\arabic{SecSecSec}}

\Sec=\z@

\def\:{\let\@sptoken= } \:  % this makes \@sptoken a space token 
\def\:{\@xifnch} \expandafter\def\: {\futurelet\@tempc\@ifnch}

\def\@ifnextchar#1#2#3{%
  \let\@tempMACe #1%
  \def\@tempMACa{#2}%
  \def\@tempMACb{#3}%
  \futurelet \@tempMACc\@ifnch%
}

\def\@ifnch{%
\ifx \@tempMACc \@sptoken%
  \let\@tempMACd\@xifnch%
\else%
  \ifx \@tempMACc \@tempMACe%
    \let\@tempMACd\@tempMACa%
  \else%
    \let\@tempMACd\@tempMACb%
  \fi%
\fi%
\@tempMACd%
}

\def\@ifstar#1#2{\@ifnextchar *{\def\@tempMACa*{#1}\@tempMACa}{#2}}

\newskip\@tempskipb

\def\addvspace#1{%
  \ifvmode\else \endgraf\fi%
  \ifdim\lastskip=\z@%
    \vskip #1\relax%
  \else%
    \@tempskipb#1\relax\@xaddvskip%
  \fi%
}

\def\@xaddvskip{%
  \ifdim\lastskip<\@tempskipb%
    \vskip-\lastskip%
    \vskip\@tempskipb\relax%
  \else%
    \ifdim\@tempskipb<\z@%
      \ifdim\lastskip<\z@ \else%
        \advance\@tempskipb\lastskip%
        \vskip-\lastskip\vskip\@tempskipb%
      \fi%
    \fi%
  \fi%
}

\newskip\@tmpSKIP

\def\addpen#1{%
  \ifvmode
    \if@nobreak
    \else
      \ifdim\lastskip=\z@
        \penalty#1\relax
      \else
        \@tmpSKIP=\lastskip
        \vskip -\lastskip
        \penalty#1\vskip\@tmpSKIP
      \fi
    \fi
  \fi
}

\newcount\@clubpen   \@clubpen=\clubpenalty
\newif\if@nobreak    \@nobreakfalse

\def\@noafterindent{%
  \global\@nobreaktrue
  \everypar{\if@nobreak
              \global\@nobreakfalse
              \clubpenalty \@M
              {\setbox\z@\lastbox}%
              \LastMac=\Nulle\relax%
            \else
              \clubpenalty \@clubpen
              \everypar{}%
            \fi}%
}

\newcount\gds@cbrk   \gds@cbrk=-300

\def\@nohdbrk{\interlinepenalty \@M\relax}

\let\@par=\par
\def\@restorepar{\def\par{\@par}}

\newif\if@endpe   \@endpefalse
 
\def\@doendpe{\@endpetrue \@nobreakfalse \LastMac=\Nulle\relax%
     \def\par{\@restorepar\everypar{}\par\@endpefalse}%
              \everypar{\setbox\z@\lastbox\everypar{}\@endpefalse}%
}

\def\section{\@ifstar{\@ssection}{\@section}}

\def\@section#1{% heading A (\section{....})
  \if@nobreak
    \everypar{}%
    \ifnum\LastMac=\Hae \addvspace{\half}\fi
  \else
    \addpen{\gds@cbrk}%
    \addvspace{\two}%
  \fi
  \bgroup
    \ninepoint\bf
    \Raggedright
    \global\advance\Sec \@ne
    \ifappendix
      \global\Eqnno=\z@ \global\SubEqnno=\z@\relax
      \def\ch@ck{#1}%
      \ifx\ch@ck\empty \def\c@lon{}\else\def\c@lon{:}\fi
      \noindent\@nohdbrk APPENDIX\ \thesection\c@lon\hskip 0.5em%
        \uppercase{#1}\par
    \else
      \noindent\@nohdbrk\thesection\hskip 1pc \uppercase{#1}\par
    \fi
    \global\SecSec=\z@
  \egroup
  \nobreak
  \vskip\half
  \nobreak
  \@noafterindent
  \LastMac=\Hae\relax
}

\def\@ssection#1{%  main section heading (\section*{....})
  \if@nobreak
    \everypar{}%
    \ifnum\LastMac=\Hae \addvspace{\half}\fi
  \else
    \addpen{\gds@cbrk}%
    \addvspace{\two}%
  \fi
  \bgroup
    \ninepoint\bf
    \Raggedright
%    \ifappendix
%      \global\Eqnno=\z@ \global\SubEqnno=\z@\relax % mh in apps dont reset
%      \noindent\@nohdbrk APPENDIX:\hskip 0.5em%
%        \uppercase{#1}\par
%    \else
    \noindent\@nohdbrk\uppercase{#1}\par
%    \fi
  \egroup
  \nobreak
  \vskip\half
  \nobreak
  \@noafterindent
  \LastMac=\Hae\relax
}

\def\subsection{\@ifstar{\@ssubsection}{\@subsection}}

\def\@subsection#1{% heading B
  \if@nobreak
    \everypar{}%
    \ifnum\LastMac=\Hae \addvspace{1pt plus 1pt minus .5pt}\fi
  \else
    \addpen{\gds@cbrk}%
    \addvspace{\onehalf}%
  \fi
  \bgroup
    \ninepoint\bf
    \Raggedright
    \global\advance\SecSec \@ne
    \noindent\@nohdbrk\thesubsection \hskip 1pc\relax #1\par
    \global\SecSecSec=\z@
  \egroup
  \nobreak
  \vskip\half
  \nobreak
  \@noafterindent
  \LastMac=\Hbe\relax
}

\def\@ssubsection#1{% heading B*
  \if@nobreak
    \everypar{}%
    \ifnum\LastMac=\Hae \addvspace{1pt plus 1pt minus .5pt}\fi
  \else
    \addpen{\gds@cbrk}%
    \addvspace{\onehalf}%
  \fi
  \bgroup
    \ninepoint\bf
    \Raggedright
    \noindent\@nohdbrk #1\par
  \egroup
  \nobreak
  \vskip\half
  \nobreak
  \@noafterindent
  \LastMac=\Hbe\relax
}

\def\subsubsection{\@ifstar{\@ssubsubsection}{\@subsubsection}}

\def\@subsubsection#1{% heading C
  \if@nobreak
    \everypar{}%
    \ifnum\LastMac=\Hbe \addvspace{1pt plus 1pt minus .5pt}\fi
  \else
    \addpen{\gds@cbrk}%
    \addvspace{\onehalf}%
  \fi
  \bgroup
    \ninepoint\it
    \Raggedright
    \global\advance\SecSecSec \@ne
    \noindent\@nohdbrk\thesubsubsection \hskip 1pc\relax #1\par
  \egroup
  \nobreak
  \vskip\half
  \nobreak
  \@noafterindent
  \LastMac=\Hce\relax
}

\def\@ssubsubsection#1{% heading C*
  \if@nobreak
    \everypar{}%
    \ifnum\LastMac=\Hbe \addvspace{1pt plus 1pt minus .5pt}\fi
  \else
    \addpen{\gds@cbrk}%
    \addvspace{\onehalf}%
  \fi
  \bgroup
    \ninepoint\it
    \Raggedright
    \noindent\@nohdbrk #1\par
  \egroup
  \nobreak
  \vskip\half
  \nobreak
  \@noafterindent
  \LastMac=\Hce\relax
}

\def\paragraph#1{% heading D
  \if@nobreak
    \everypar{}%
  \else
    \addpen{\gds@cbrk}%
    \addvspace{\one}%
  \fi%
  \bgroup%
    \ninepoint\it
    \noindent #1\ \nobreak%
  \egroup
  \LastMac=\Hde\relax
  \ignorespaces
}

% Appendix

\newif\ifappendix

\def\appendix{%
  \global\appendixtrue
  \def\thesection{\Alph{Sec}}%
  \def\thesubsection{\thesection\arabic{SecSec}}%
  \def\theeq{\thesection\arabic{Eqnno}}%
  \Sec=\z@ \SecSec=\z@ \SecSecSec=\z@ \Eqnno=\z@ \SubEqnno=\z@\relax
}

% Text

 % provided for backward compatibility

% Lists

\def\beginlist{%
  \par\if@nobreak \else\addvspace{\half}\fi%
  \bgroup%
    \ninepoint
    \let\item=\list@item%
}

\def\list@item{%
  \par\noindent\hskip 1em\relax%
  \ignorespaces%
}

\def\endlist{\par\egroup\addvspace{\half}\@doendpe}

% References

\def\beginrefs{%
  \par
  \bgroup
    \eightpoint
    \Fullout
    \let\bibitem=\bib@item
}

\def\bib@item{%
  \par\parindent=1.5em\Hang{1.5em}{1}%
  \everypar={\Hang{1.5em}{1}\ignorespaces}%
  \noindent\ignorespaces
}

\def\endrefs{\par\egroup\@doendpe}

% Page heads

\newtoks\CatchLine

\def\@journal{Mon.\ Not.\ R.\ Astron.\ Soc.\ }  % The journal title string
\def\@pubyear{1994}        % Assign a default publication year
\def\@pagerange{000--000}  % Assign a default page-range
\def\@volume{000}          % Assign a default volume number
\def\@microfiche{}         %

\def\pubyear#1{\gdef\@pubyear{#1}\@makecatchline}
\def\pagerange#1{\gdef\@pagerange{#1}\@makecatchline}
\def\volume#1{\gdef\@volume{#1}\@makecatchline}
\def\microfiche#1{\gdef\@microfiche{and Microfiche\ #1}\@makecatchline}

\def\@makecatchline{%
  \global\CatchLine{%
    {\rm \@journal {\bf \@volume},\ \@pagerange\ (\@pubyear)\ \@microfiche}}%
}

\@makecatchline % Assign a catchline, using the above defaults

\newtoks\LeftHeader
\def\shortauthor#1{% left page head
  \global\LeftHeader{#1}%
}

\newtoks\RightHeader
\def\shorttitle#1{% right page head
  \global\RightHeader{#1}%
}

\def\PageHead{% recto/verso running heads
  \begingroup
    \ifsp@page
      \csname ps@\sp@type\endcsname
    \fi
    \ifodd\pageno
      \let\the@head=\@oddhead
    \else
      \let\the@head=\@evenhead
    \fi
    \vbox to \z@{\vskip-22.5\p@%
      \hbox to \PageWidth{\vbox to8.5\p@{}%
        \the@head
      }%
    \vss}%
  \endgroup
  \nointerlineskip
}

\gdef\PageFoot{%
  \nointerlineskip%
  \begingroup
  \ifsp@page
    \csname ps@\sp@type\endcsname
    \global\sp@pagefalse
  \fi
  \vbox to 22pt{\vfil%
    \hbox to \PageWidth{%
      \eightpoint\strut\noindent
      \ifodd\pageno
        \@oddfoot
      \else
        \@evenfoot
      \fi
    }%
  }%
  \endgroup
}

\def\today{%
  \number\day\space
  \ifcase\month\or January\or February\or March\or April\or May\or June\or
    July\or August\or September\or October\or November\or December\fi
  \space\number\year%
}

\def\authorcomment#1{%
  \gdef\PageFoot{%
    \nointerlineskip%
    \vbox to 20pt{\vfil%
      \hbox to \PageWidth{\elevenpoint\noindent \hfil #1 \hfil}}%
  }%
}

% Plate pages

\newif\ifplate@page
\newbox\plt@box

\def\beginplatepage{%
  \let\plate=\plate@head
  \let\caption=\fig@caption
  \global\setbox\plt@box=\vbox\bgroup
  \TEMPDIMEN=\PageWidth % For \fig@caption test
  \hsize=\PageWidth\relax
}

\def\endplatepage{\par\egroup\global\plate@pagetrue}
\def\plate@head#1{\gdef\plt@cap{#1}}

% Letters option

\def\letters{%
  \gdef\folio{\ifnum\pageno<\z@ L\romannumeral-\pageno
    \else L\number\pageno \fi}%
}

% Math setup

% The standard math indentation
\newdimen\mathindent

\global\mathindent=\z@
\global\everydisplay{\global\@dspwd=\displaywidth\displaysetup}

% New versions of \displaylines, \eqalign, \eqalignno for
% when non-centered math is in use.

\def\@displaylines#1{% (for non-centered math)
  {}$\displ@y\hbox{\vbox{\halign{$\@lign\hfil\displaystyle##\hfil$\crcr
  #1\crcr}}}${}%
}

\def\@eqalign#1{\null\vcenter{\openup\jot\m@th% (for non-centered math)
  \ialign{\strut\hfil$\displaystyle{##}$&$\displaystyle{{}##}$\hfil
      \crcr#1\crcr}}%
}

\def\@eqalignno#1{% (for non-centered math)
  \global\advance\@dspwd by -\mathindent%
  {}$\displ@y\hbox{\vbox{\halign to\@dspwd%
  {\hfil$\@lign\displaystyle{##}$\tabskip\z@skip
  &$\@lign\displaystyle{{}##}$\hfil\tabskip\centering
  &\llap{$\@lign##$}\tabskip\z@skip\crcr
  #1\crcr}}}${}%
}

% When equations are flushleft ensure, that \displaylines,
% \eqalign, \eqalignno and \leqalignno point to the new versions of
% the macros. Also make \leqalignno act like \eqalignno, otherwise the
% equation text would `crash' into the equation number.

\global\let\displaylines=\@displaylines
\global\let\eqalign=\@eqalign
\global\let\eqalignno=\@eqalignno
\global\let\leqalignno=\@eqalignno

\newdimen\@dspwd   \@dspwd=\z@
\newif\if@eqno
\newif\if@leqno
\newtoks\@eqn
\newtoks\@eq

\def\displaysetup#1$${\displaytest#1\eqno\eqno\displaytest}

\def\displaytest#1\eqno#2\eqno#3\displaytest{%
 \if!#3!\ldisplaytest#1\leqno\leqno\ldisplaytest
 \else\@eqnotrue\@leqnofalse\@eqn={#2}\@eq={#1}\fi
 \generaldisplay$$}

\def\ldisplaytest#1\leqno#2\leqno#3\ldisplaytest{%
\@eq={#1}%
 \if!#3!\@eqnofalse\else\@eqnotrue\@leqnotrue
  \@eqn={#2}\fi}

\def\generaldisplay{%
  \if@eqno
    \if@leqno
      \hbox to \displaywidth{\noindent
        \rlap{$\displaystyle\the\@eqn$}%
        \hskip\mathindent$\displaystyle\the\@eq$\hfil}%
    \else
      \hbox to \displaywidth{\noindent
        \hskip\mathindent
        $\displaystyle\the\@eq$\hfil$\displaystyle\the\@eqn$}%
    \fi
  \else
    \hbox to \displaywidth{\noindent
      \hskip\mathindent$\displaystyle\the\@eq$\hfil}%
  \fi
}

% Finishing notice

\def\@notice{%
  \par\Two%
  \noindent{\b@ls{11pt}\ninerm This paper has been produced using the
    Royal Astronomical Society/Blackwell Science \TeX\ macros.\par}%
}

% redefine \bye to output our identification notice :
\outer\def\bye{\@notice\par\vfill\supereject\end}

% define a sign on :

\def\start@mess{%
  Monthly notices of the RAS journal style (\@typeface)\space
    v\@version,\space \@verdate.%
}

\everyjob{\Warn{\start@mess}}

% Two-column macros

%--------------------------------------------------------%
%                     INITIALISATION                     %
%--------------------------------------------------------%

\newif\if@debug \@debugfalse  %  when false, only warnings displayed

\def\Print#1{\if@debug\immediate\write16{#1}\else \fi}
\def\Warn#1{\immediate\write16{#1}}
\def\wlog#1{}

\newcount\Iteration % temporary loop counter

\def\Single{0} \def\Double{1}                 % ItemSPAN's
\def\Figure{0} \def\Table{1}                  % ItemTYPE's

\def\InStack{0}  % ItemSTATUS
\def\InZoneA{1}
\def\InZoneB{2}
\def\InZoneC{3}

\newcount\TEMPCOUNT % temporary count register
\newdimen\TEMPDIMEN % temporary dimen register
\newbox\TEMPBOX     % temporary box register
\newbox\VOIDBOX     % a box which is permenately void

\newcount\LengthOfStack % number of items currently in stack
\newcount\MaxItems      % maximum number of items allowed in stack
\newcount\StackPointer
\newcount\Point         % used in calculation for generating the
                        % physical address of an item in the stack.
\newcount\NextFigure    % number of next figure to be output
\newcount\NextTable     % number of next table to be output
\newcount\NextItem      % Next item to consider by order in stack

\newcount\StatusStack   % set to point to top of STATUS stack
\newcount\NumStack      % set to point to top of NUMBER stack
\newcount\TypeStack     % set to point to top of TYPE stack
\newcount\SpanStack     % set to point to top of SPAN stack
\newcount\BoxStack      % set to point to top of BOX stack

\newcount\ItemSTATUS    % status of present item
\newcount\ItemNUMBER    % number of present item
\newcount\ItemTYPE      % type of present item
\newcount\ItemSPAN      % span of present item
\newbox\ItemBOX         % box of present item
\newdimen\ItemSIZE      % size of present item
                        % (calculated by GetItemBOX)

\newdimen\PageHeight    % vertical measure of full page
\newdimen\TextLeading   % distance between baselines of body text
\newdimen\Feathering    % amount of interline stretch
\newcount\LinesPerPage  % height of page in text lines
\newdimen\ColumnWidth   % width of 1 column of text
\newdimen\ColumnGap     % size of gap between columns
\newdimen\PageWidth     % = \ColumnWidth * 2 + \ColumnGap
\newdimen\BodgeHeight   % Bodge to align figures and tables with text
\newcount\Leading       % Set to same as \TextLeading above

\newdimen\ZoneBSize  % size of items in ZoneB
\newdimen\TextSize   % size of text in ZoneB
\newbox\ZoneABOX     % contains Zone A material
\newbox\ZoneBBOX     % contains Zone B material
\newbox\ZoneCBOX     % contains Zone C material

\newif\ifFirstSingleItem
\newif\ifFirstZoneA
\newif\ifMakePageInComplete
\newif\ifMoreFigures \MoreFiguresfalse % set true in join stack
\newif\ifMoreTables  \MoreTablesfalse  % set true in join stack

\newif\ifFigInZoneB % used to determine in which zone an item
\newif\ifFigInZoneC % will be placed based on what is in other
\newif\ifTabInZoneB % zones already for a given page.
\newif\ifTabInZoneC

\newif\ifZoneAFullPage

\newbox\MidBOX    % = LeftBOX+gap+RightBOX
\newbox\LeftBOX
\newbox\RightBOX
\newbox\PageBOX   % complete made-up page

\newif\ifLeftCOL  % flags first pass through output routine
\LeftCOLtrue

\newdimen\ZoneBAdjust

\newcount\ItemFits
\def\Yes{1}
\def\No{2}

% Setup file.

\MaxItems=15
\NextFigure=\z@        % used for article opening
\NextTable=\@ne

\BodgeHeight=6pt
\TextLeading=11pt    % baselineskip of body text
\Leading=11
\Feathering=\z@      % amount of interline stretch
\LinesPerPage=61     % number of text lines per full page -1
\topskip=\TextLeading
\ColumnWidth=20pc    % width of text columns
\ColumnGap=2pc       % gap between columns

\newskip\ItemSepamount  % space between floats
\ItemSepamount=\TextLeading plus \TextLeading minus 4pt

\parskip=\z@ plus .1pt
\parindent=18pt
\widowpenalty=\z@
\clubpenalty=10000
\tolerance=1500
\hbadness=1500
\abovedisplayskip=6pt plus 2pt minus 1pt
\belowdisplayskip=6pt plus 2pt minus 1pt
\abovedisplayshortskip=6pt plus 2pt minus 1pt
\belowdisplayshortskip=6pt plus 2pt minus 1pt

\frenchspacing

\ninepoint % start main text size

\PageHeight=682pt
\PageWidth=2\ColumnWidth
\advance\PageWidth by \ColumnGap

\pagestyle{headings}

%--------------------------------------------------------%
%                         STACKS                         %
%--------------------------------------------------------%

% THE ITEM STACK
% The item stack contains contains figures and tables
% in the order in which they appear in the article source
% code.

% allocate stack space

\newcount\DUMMY \StatusStack=\allocationnumber
\newcount\DUMMY \newcount\DUMMY \newcount\DUMMY 
\newcount\DUMMY \newcount\DUMMY \newcount\DUMMY 
\newcount\DUMMY \newcount\DUMMY \newcount\DUMMY
\newcount\DUMMY \newcount\DUMMY \newcount\DUMMY 
\newcount\DUMMY \newcount\DUMMY \newcount\DUMMY

\newcount\DUMMY \NumStack=\allocationnumber
\newcount\DUMMY \newcount\DUMMY \newcount\DUMMY 
\newcount\DUMMY \newcount\DUMMY \newcount\DUMMY 
\newcount\DUMMY \newcount\DUMMY \newcount\DUMMY 
\newcount\DUMMY \newcount\DUMMY \newcount\DUMMY 
\newcount\DUMMY \newcount\DUMMY \newcount\DUMMY

\newcount\DUMMY \TypeStack=\allocationnumber
\newcount\DUMMY \newcount\DUMMY \newcount\DUMMY 
\newcount\DUMMY \newcount\DUMMY \newcount\DUMMY 
\newcount\DUMMY \newcount\DUMMY \newcount\DUMMY 
\newcount\DUMMY \newcount\DUMMY \newcount\DUMMY 
\newcount\DUMMY \newcount\DUMMY \newcount\DUMMY

\newcount\DUMMY \SpanStack=\allocationnumber
\newcount\DUMMY \newcount\DUMMY \newcount\DUMMY 
\newcount\DUMMY \newcount\DUMMY \newcount\DUMMY 
\newcount\DUMMY \newcount\DUMMY \newcount\DUMMY 
\newcount\DUMMY \newcount\DUMMY \newcount\DUMMY 
\newcount\DUMMY \newcount\DUMMY \newcount\DUMMY

\newbox\DUMMY   \BoxStack=\allocationnumber
\newbox\DUMMY   \newbox\DUMMY \newbox\DUMMY 
\newbox\DUMMY   \newbox\DUMMY \newbox\DUMMY 
\newbox\DUMMY   \newbox\DUMMY \newbox\DUMMY 
\newbox\DUMMY   \newbox\DUMMY \newbox\DUMMY 
\newbox\DUMMY   \newbox\DUMMY \newbox\DUMMY

\def\wlog{\immediate\write\m@ne}

% \GetItemSTATUS, \GetItemNUMBER, \GetItemTYPE, \GetItemSPAN,
% \GetItemBox 
% are used to get details of a particular item from the item
% stack. The argument to each of these is the items position
% in the stack (usually \StackPointer)...not the items number.

\def\GetItemAll#1{%
 \GetItemSTATUS{#1}
 \GetItemNUMBER{#1}
 \GetItemTYPE{#1}
 \GetItemSPAN{#1}
 \GetItemBOX{#1}
}

% Note: \LeaveStack uses this routine. Do not destroy \Point
\def\GetItemSTATUS#1{%
 \Point=\StatusStack
 \advance\Point by #1
 \global\ItemSTATUS=\count\Point
}

% Note: \LeaveStack uses this routine. Do not destroy \Point
\def\GetItemNUMBER#1{%
 \Point=\NumStack
 \advance\Point by #1
 \global\ItemNUMBER=\count\Point
}

% Note: \LeaveStack uses this routine. Do not destroy \Point
\def\GetItemTYPE#1{%
 \Point=\TypeStack
 \advance\Point by #1
 \global\ItemTYPE=\count\Point
}

% Note: \LeaveStack uses this routine. Do not destroy \Point
\def\GetItemSPAN#1{%
 \Point\SpanStack
 \advance\Point by #1
 \global\ItemSPAN=\count\Point
}

% Note: \LeaveStack uses this routine. Do not destroy \Point
\def\GetItemBOX#1{%
 \Point=\BoxStack
 \advance\Point by #1
 \global\setbox\ItemBOX=\vbox{\copy\Point}
 \global\ItemSIZE=\ht\ItemBOX
 \global\advance\ItemSIZE by \dp\ItemBOX
 \TEMPCOUNT=\ItemSIZE
 \divide\TEMPCOUNT by \Leading
 \divide\TEMPCOUNT by 65536
 \advance\TEMPCOUNT \@ne
 \ItemSIZE=\TEMPCOUNT pt
 \global\multiply\ItemSIZE by \Leading
}

% item joins stack

\def\JoinStack{%
 \ifnum\LengthOfStack=\MaxItems % stack is full of items
  \Warn{WARNING: Stack is full...some items will be lost!}
 \else
  \Point=\StatusStack
  \advance\Point by \LengthOfStack
  \global\count\Point=\ItemSTATUS
  \Point=\NumStack
  \advance\Point by \LengthOfStack
  \global\count\Point=\ItemNUMBER
  \Point=\TypeStack
  \advance\Point by \LengthOfStack
  \global\count\Point=\ItemTYPE
  \Point\SpanStack
  \advance\Point by \LengthOfStack
  \global\count\Point=\ItemSPAN
  \Point=\BoxStack
  \advance\Point by \LengthOfStack
  \global\setbox\Point=\vbox{\copy\ItemBOX}
  \global\advance\LengthOfStack \@ne
  \ifnum\ItemTYPE=\Figure % used in \MakePage
   \global\MoreFigurestrue
  \else
   \global\MoreTablestrue
  \fi
 \fi
}

% item leaves stack
% #1=physical position of the item to be removed

\def\LeaveStack#1{%
 {\Iteration=#1
 \loop
 \ifnum\Iteration<\LengthOfStack
  \advance\Iteration \@ne
  \GetItemSTATUS{\Iteration}
   \advance\Point by \m@ne
   \global\count\Point=\ItemSTATUS
  \GetItemNUMBER{\Iteration}
   \advance\Point by \m@ne
   \global\count\Point=\ItemNUMBER
  \GetItemTYPE{\Iteration}
   \advance\Point by \m@ne
   \global\count\Point=\ItemTYPE
  \GetItemSPAN{\Iteration}
   \advance\Point by \m@ne
   \global\count\Point=\ItemSPAN
  \GetItemBOX{\Iteration}
   \advance\Point by \m@ne
   \global\setbox\Point=\vbox{\copy\ItemBOX}
 \repeat}
 \global\advance\LengthOfStack by \m@ne
}

% clean stack
% This routine scans through the stack and removes anything
% that does not have STATUS=\InStack.

\newif\ifStackNotClean

\def\CleanStack{%
 \StackNotCleantrue
 {\Iteration=\z@
  \loop
   \ifStackNotClean
    \GetItemSTATUS{\Iteration}
    \ifnum\ItemSTATUS=\InStack
     \advance\Iteration \@ne
     \else
      \LeaveStack{\Iteration}
    \fi
   \ifnum\LengthOfStack<\Iteration
    \StackNotCleanfalse
   \fi
 \repeat}
}

% Find item.
% This macro searches from the top to the bottom of the
% stack for an item of a specified type and number.
% #1=type, #2=number
% If the specified item is found, then \StackPointer is set
% to point to it, else \StackPointer=-1.
% This routine is used to find the next figure or table
% by number.

\def\FindItem#1#2{%
 \global\StackPointer=\m@ne % assume item isn't in stack for now
 {\Iteration=\z@
  \loop
  \ifnum\Iteration<\LengthOfStack
   \GetItemSTATUS{\Iteration}
   \ifnum\ItemSTATUS=\InStack
    \GetItemTYPE{\Iteration}
    \ifnum\ItemTYPE=#1
     \GetItemNUMBER{\Iteration}
     \ifnum\ItemNUMBER=#2
      \global\StackPointer=\Iteration
      \Iteration=\LengthOfStack % terminate loop
     \fi
    \fi
   \fi
  \advance\Iteration \@ne
 \repeat}
}

% Find next type
% This macro searches from the top to the bottom of the stack
% looking for the first item which has STATUS=\InStack.
% If it is a figure then a figure is what will be considered
% next by \MakePage else table.

\def\FindNext{%
 \global\StackPointer=\m@ne % assume stack is empty for now
 {\Iteration=\z@
  \loop
  \ifnum\Iteration<\LengthOfStack
   \GetItemSTATUS{\Iteration}
   \ifnum\ItemSTATUS=\InStack
    \GetItemTYPE{\Iteration}
   \ifnum\ItemTYPE=\Figure
    \ifMoreFigures
      \global\NextItem=\Figure
      \global\StackPointer=\Iteration
      \Iteration=\LengthOfStack % terminate loop
    \fi
   \fi
   \ifnum\ItemTYPE=\Table
    \ifMoreTables
      \global\NextItem=\Table
      \global\StackPointer=\Iteration
      \Iteration=\LengthOfStack % terminate loop
    \fi
   \fi
  \fi
  \advance\Iteration \@ne
 \repeat}
}

% Change status
% Macro to change the status of a specified item in stack.
% #1=item, #2=new status

\def\ChangeStatus#1#2{%
 \Point=\StatusStack
 \advance\Point by #1
 \global\count\Point=#2
}

%--------------------------------------------------------%
%                       MAKEPAGE                         %
%--------------------------------------------------------%

% This macro is called at the start of every new page
% including the first. It scans through the stack picking
% out items which should be placed on this page. It then
% leaves space for the items to be placed later. The routine
% terminates when either there is no room on the page to
% fit the next figure or table, or there are no more items
% in the stack.

\def\Zone{\InZoneA}

\ZoneBAdjust=\z@

\def\MakePage{% allocate space on this page for stack items
 \global\ZoneBSize=\PageHeight
 \global\TextSize=\ZoneBSize
 \global\ZoneAFullPagefalse
 \global\topskip=\TextLeading
 \MakePageInCompletetrue
 \MoreFigurestrue
 \MoreTablestrue
 \FigInZoneBfalse
 \FigInZoneCfalse
 \TabInZoneBfalse
 \TabInZoneCfalse
 \global\FirstSingleItemtrue
 \global\FirstZoneAtrue
 \global\setbox\ZoneABOX=\box\VOIDBOX
 \global\setbox\ZoneBBOX=\box\VOIDBOX
 \global\setbox\ZoneCBOX=\box\VOIDBOX
 \loop
  \ifMakePageInComplete
 \FindNext
 \ifnum\StackPointer=\m@ne
  \NextItem=\m@ne
  \MoreFiguresfalse
  \MoreTablesfalse
 \fi
 \ifnum\NextItem=\Figure
   \FindItem{\Figure}{\NextFigure}
   \ifnum\StackPointer=\m@ne \global\MoreFiguresfalse
   \else
    \GetItemSPAN{\StackPointer}
    \ifnum\ItemSPAN=\Single \def\Zone{\InZoneB}\relax
     \ifFigInZoneC \global\MoreFiguresfalse\fi
    \else
     \def\Zone{\InZoneA}
     \ifFigInZoneB \def\Zone{\InZoneC}\fi
    \fi
   \fi
   \ifMoreFigures\Print{}\FigureItems\fi
 \fi
\ifnum\NextItem=\Table
   \FindItem{\Table}{\NextTable}
   \ifnum\StackPointer=\m@ne \global\MoreTablesfalse
   \else
    \GetItemSPAN{\StackPointer}
    \ifnum\ItemSPAN=\Single\relax
     \ifTabInZoneC \global\MoreTablesfalse\fi
    \else
     \def\Zone{\InZoneA}
     \ifTabInZoneB \def\Zone{\InZoneC}\fi
    \fi
   \fi
   \ifMoreTables\Print{}\TableItems\fi
 \fi
   \MakePageInCompletefalse % assume page is complete
   \ifMoreFigures\MakePageInCompletetrue\fi
   \ifMoreTables\MakePageInCompletetrue\fi
 \repeat
%\Print{TextSize=\the\TextSize}
%\Print{ZoneBSize=\the\ZoneBSize}
 \ifZoneAFullPage
  \global\TextSize=\z@
  \global\ZoneBSize=\z@
  \global\vsize=\z@\relax
  \global\topskip=\z@\relax
  \vbox to \z@{\vss}
  \eject
 \else
 \global\advance\ZoneBSize by -\ZoneBAdjust
 \global\vsize=\ZoneBSize
 \global\hsize=\ColumnWidth
 \global\ZoneBAdjust=\z@
 \ifdim\TextSize<23pt
 \Warn{}
 \Warn{* Making column fall short: TextSize=\the\TextSize *}
 \vskip-\lastskip\eject\fi
 \fi
}

\def\MakeRightCol{% allocate space for the right column of text
 \global\TextSize=\ZoneBSize
 \MakePageInCompletetrue
 \MoreFigurestrue
 \MoreTablestrue
 \global\FirstSingleItemtrue
 \global\setbox\ZoneBBOX=\box\VOIDBOX
 \def\Zone{\InZoneB}
 \loop
  \ifMakePageInComplete
 \FindNext
 \ifnum\StackPointer=\m@ne
  \NextItem=\m@ne
  \MoreFiguresfalse
  \MoreTablesfalse
 \fi
 \ifnum\NextItem=\Figure
   \FindItem{\Figure}{\NextFigure}
   \ifnum\StackPointer=\m@ne \MoreFiguresfalse
   \else
    \GetItemSPAN{\StackPointer}
    \ifnum\ItemSPAN=\Double\relax
     \MoreFiguresfalse\fi
   \fi
   \ifMoreFigures\Print{}\FigureItems\fi
 \fi
 \ifnum\NextItem=\Table
   \FindItem{\Table}{\NextTable}
   \ifnum\StackPointer=\m@ne \MoreTablesfalse
   \else
    \GetItemSPAN{\StackPointer}
    \ifnum\ItemSPAN=\Double\relax
     \MoreTablesfalse\fi
   \fi
   \ifMoreTables\Print{}\TableItems\fi
 \fi
   \MakePageInCompletefalse % assume page is complete
   \ifMoreFigures\MakePageInCompletetrue\fi
   \ifMoreTables\MakePageInCompletetrue\fi
 \repeat
 \ifZoneAFullPage
  \global\TextSize=\z@
  \global\ZoneBSize=\z@
  \global\vsize=\z@\relax
  \global\topskip=\z@\relax
  \vbox to \z@{\vss}
  \eject
 \else
 \global\vsize=\ZoneBSize
 \global\hsize=\ColumnWidth
 \ifdim\TextSize<23pt
 \Warn{}
 \Warn{* Making column fall short: TextSize=\the\TextSize *}
 \vskip-\lastskip\eject\fi
\fi
}

\def\FigureItems{% Stack pointer points to next figure
 \Print{Considering...}
 \ShowItem{\StackPointer}
 \GetItemBOX{\StackPointer} % auto calculates ItemSIZE
 \GetItemSPAN{\StackPointer}
  \CheckFitInZone % check to see if item fits
  \ifnum\ItemFits=\Yes
   \ifnum\ItemSPAN=\Single
     \ChangeStatus{\StackPointer}{\InZoneB} % flag to be output
     \global\FigInZoneBtrue
     \ifFirstSingleItem
      \hbox{}\vskip-\BodgeHeight
     \global\advance\ItemSIZE by \TextLeading
     \fi
     \unvbox\ItemBOX\ItemSep
     \global\FirstSingleItemfalse
     \global\advance\TextSize by -\ItemSIZE% allocate space
     \global\advance\TextSize by -\TextLeading
   \else
    \ifFirstZoneA
     \global\advance\ItemSIZE by \TextLeading
     \global\FirstZoneAfalse\fi
    \global\advance\TextSize by -\ItemSIZE
    \global\advance\TextSize by -\TextLeading
    \global\advance\ZoneBSize by -\ItemSIZE
    \global\advance\ZoneBSize by -\TextLeading
    \ifFigInZoneB\relax
     \else
     \ifdim\TextSize<3\TextLeading
     \global\ZoneAFullPagetrue
     \fi
    \fi
    \ChangeStatus{\StackPointer}{\Zone}
    \ifnum\Zone=\InZoneC \global\FigInZoneCtrue\fi
  \fi
   \Print{TextSize=\the\TextSize}
   \Print{ZoneBSize=\the\ZoneBSize}
  \global\advance\NextFigure \@ne
   \Print{This figure has been placed.}
  \else
   \Print{No space available for this figure...holding over.}
   \Print{}
   \global\MoreFiguresfalse
  \fi
}

\def\TableItems{% Stack pointer points to next table
 \Print{Considering...}
 \ShowItem{\StackPointer}
 \GetItemBOX{\StackPointer} % auto calculates ItemSIZE
 \GetItemSPAN{\StackPointer}
  \CheckFitInZone % check to see of item fits in Zone
  \ifnum\ItemFits=\Yes
   \ifnum\ItemSPAN=\Single
    \ChangeStatus{\StackPointer}{\InZoneB}
     \global\TabInZoneBtrue
     \ifFirstSingleItem
      \hbox{}\vskip-\BodgeHeight
     \global\advance\ItemSIZE by \TextLeading
     \fi
     \unvbox\ItemBOX\ItemSep
     \global\FirstSingleItemfalse
     \global\advance\TextSize by -\ItemSIZE
     \global\advance\TextSize by -\TextLeading
   \else
    \ifFirstZoneA
    \global\advance\ItemSIZE by \TextLeading
    \global\FirstZoneAfalse\fi
    \global\advance\TextSize by -\ItemSIZE
    \global\advance\TextSize by -\TextLeading
    \global\advance\ZoneBSize by -\ItemSIZE
    \global\advance\ZoneBSize by -\TextLeading
    \ifFigInZoneB\relax
     \else
     \ifdim\TextSize<3\TextLeading
     \global\ZoneAFullPagetrue
     \fi
    \fi
    \ChangeStatus{\StackPointer}{\Zone}
    \ifnum\Zone=\InZoneC \global\TabInZoneCtrue\fi
   \fi
%   \Print{TextSize=\the\TextSize}
%   \Print{ZoneBSize=\the\ZoneBSize}
  \global\advance\NextTable \@ne
   \Print{This table has been placed.}
  \else
  \Print{No space available for this table...holding over.}
   \Print{}
   \global\MoreTablesfalse
  \fi
}

% These macros check to see if an item of ItemSIZE will
% fit in a particular zone. If it will, then ItemFits
% will be set true else false.

\def\CheckFitInZone{%
{\advance\TextSize by -\ItemSIZE
 \advance\TextSize by -\TextLeading
 \ifFirstSingleItem
  \advance\TextSize by \TextLeading
 \fi
 \ifnum\Zone=\InZoneA\relax
  \else \advance\TextSize by -\ZoneBAdjust
 \fi
 \ifdim\TextSize<3\TextLeading \global\ItemFits=\No
 \else \global\ItemFits=\Yes\fi}
}

\def\BeginOpening{%
  % start 9pt a.s.a.p. so that \New.. commands get a chance to init.
  \ninepoint
  \thispagestyle{titlepage}%
  \global\setbox\ItemBOX=\vbox\bgroup%
    \hsize=\PageWidth%
    \hrule height \z@
    \ifsinglecol\vskip 6pt\fi % Bodge, to get same pos. as two-column!
}

\let\begintopmatter=\BeginOpening  %  alias for \BeginOpening

\def\EndOpening{%
  \One%  1 line fixed space below opening
  \egroup
  \ifsinglecol
    \box\ItemBOX%
    \vskip\TextLeading plus 2\TextLeading% var. space: min 1, max 3 lines
    \@noafterindent
  \else
    \ItemNUMBER=\z@%
    \ItemTYPE=\Figure
    \ItemSPAN=\Double
    \ItemSTATUS=\InStack
    \JoinStack
  \fi
}

% Figures

\newif\if@here  \@herefalse

\def\no@float{\global\@heretrue}
\let\nofloat=\relax % only enabled for one column

\def\beginfigure{%
  \@ifstar{\global\@dfloattrue \@bfigure}{\global\@dfloatfalse \@bfigure}%
}

\def\@bfigure#1{%
  \par
  \if@dfloat
    \ItemSPAN=\Double
    \TEMPDIMEN=\PageWidth
  \else
    \ItemSPAN=\Single
    \TEMPDIMEN=\ColumnWidth
  \fi
  \ifsinglecol
    \TEMPDIMEN=\PageWidth
  \else
    \ItemSTATUS=\InStack
    \ItemNUMBER=#1%
    \ItemTYPE=\Figure
  \fi
  \bgroup
    \hsize=\TEMPDIMEN
    \global\setbox\ItemBOX=\vbox\bgroup
      \eightpoint\nostb@ls{10pt}%
      \let\caption=\fig@caption
      \ifsinglecol \let\nofloat=\no@float\fi
}

\def\fig@caption#1{%
  \vskip 5.5pt plus 6pt%
  \bgroup % grouping and size change needed for plate pages
    \eightpoint\nostb@ls{10pt}%
    \setbox\TEMPBOX=\hbox{#1}%
    \ifdim\wd\TEMPBOX>\TEMPDIMEN
      \noindent \unhbox\TEMPBOX\par
    \else
      \hbox to \hsize{\hfil\unhbox\TEMPBOX\hfil}%
    \fi
  \egroup
}

\def\endfigure{%
  \par\egroup % end \vbox
  \egroup
  \ifsinglecol
    \if@here \midinsert\global\@herefalse\else \topinsert\fi
      \unvbox\ItemBOX
    \endinsert
  \else
    \JoinStack
    \Print{Processing source for figure \the\ItemNUMBER}%
  \fi
}

% Tables

\newbox\tab@cap@box
\def\tab@caption#1{\global\setbox\tab@cap@box=\hbox{#1\par}}

\newtoks\tab@txt@toks
\long\def\tab@txt#1{\global\tab@txt@toks={#1}\global\table@txttrue}

\newif\iftable@txt  \table@txtfalse
\newif\if@dfloat    \@dfloatfalse

\def\begintable{%
  \@ifstar{\global\@dfloattrue \@btable}{\global\@dfloatfalse \@btable}%
}

\def\@btable#1{%
  \par
  \if@dfloat
    \ItemSPAN=\Double
    \TEMPDIMEN=\PageWidth
  \else
    \ItemSPAN=\Single
    \TEMPDIMEN=\ColumnWidth
  \fi
  \ifsinglecol
    \TEMPDIMEN=\PageWidth
  \else
    \ItemSTATUS=\InStack
    \ItemNUMBER=#1%
    \ItemTYPE=\Table
  \fi
  \bgroup
    \eightpoint\nostb@ls{10pt}%
    \global\setbox\ItemBOX=\vbox\bgroup
      \let\caption=\tab@caption
      \let\tabletext=\tab@txt
      \ifsinglecol \let\nofloat=\no@float\fi
}

\def\endtable{%
  \par\egroup % end \vbox
  \egroup
  \setbox\TEMPBOX=\hbox to \TEMPDIMEN{%
    \eightpoint\nostb@ls{10pt}%
    \hss
    \vbox{%
      \hsize=\wd\ItemBOX
      \ifvoid\tab@cap@box
      \else
        \noindent\unhbox\tab@cap@box
        \vskip 5.5pt plus 6pt%
      \fi
      \box\ItemBOX
      \iftable@txt
        \vskip 10pt%
        \noindent\the\tab@txt@toks
        \global\table@txtfalse
      \fi
    }%
    \hss
  }%
  \ifsinglecol
    \if@here \midinsert\global\@herefalse\else \topinsert\fi
      \box\TEMPBOX
    \endinsert
  \else
    \global\setbox\ItemBOX=\box\TEMPBOX
    \JoinStack
    \Print{Processing source for table \the\ItemNUMBER}%
  \fi
}

\def\UnloadZoneA{%
\FirstZoneAtrue
 \Iteration=\z@
  \loop
   \ifnum\Iteration<\LengthOfStack
    \GetItemSTATUS{\Iteration}
    \ifnum\ItemSTATUS=\InZoneA
     \GetItemBOX{\Iteration}
     \ifFirstZoneA \vbox to \BodgeHeight{\vfil}%
     \FirstZoneAfalse\fi
     \unvbox\ItemBOX\ItemSep
     \LeaveStack{\Iteration}
     \else
     \advance\Iteration \@ne
   \fi
 \repeat
}

\def\UnloadZoneC{%
\Iteration=\z@
  \loop
   \ifnum\Iteration<\LengthOfStack
    \GetItemSTATUS{\Iteration}
    \ifnum\ItemSTATUS=\InZoneC
     \GetItemBOX{\Iteration}
     \ItemSep\unvbox\ItemBOX
     \LeaveStack{\Iteration}
     \else
     \advance\Iteration \@ne
   \fi
 \repeat
}

%--------------------------------------------------------%
%                         DIAGNOSTICS                    %
%--------------------------------------------------------%

\def\ShowItem#1{% Show details of on item entry in stack
  {\GetItemAll{#1}
  \Print{\the#1:
  {TYPE=\ifnum\ItemTYPE=\Figure Figure\else Table\fi}
  {NUMBER=\the\ItemNUMBER}
  {SPAN=\ifnum\ItemSPAN=\Single Single\else Double\fi}
  {SIZE=\the\ItemSIZE}}}
}

\def\ShowStack{% 
 \Print{}
 \Print{LengthOfStack = \the\LengthOfStack}
 \ifnum\LengthOfStack=\z@ \Print{Stack is empty}\fi
 \Iteration=\z@
 \loop
 \ifnum\Iteration<\LengthOfStack
  \ShowItem{\Iteration}
  \advance\Iteration \@ne
 \repeat
}

\def\B#1#2{%
\hbox{\vrule\kern-0.4pt\vbox to #2{%
\hrule width #1\vfill\hrule}\kern-0.4pt\vrule}
}

%-------------------------------------------------------%
%             SINGLE COLUMN OUTPUT ROUTINE              %
%-------------------------------------------------------%

\newif\ifsinglecol   \singlecolfalse

\def\onecolumn{%
  \global\output={\singlecoloutput}%
  \global\hsize=\PageWidth
  \global\vsize=\PageHeight
  \global\ColumnWidth=\hsize
  \global\TextLeading=12pt
  \global\Leading=12
  \global\singlecoltrue
  \global\let\onecolumn=\relax%         stop them using \onecolumn again
  \global\let\footnote=\sing@footnote%  enable footnotes
  \global\let\vfootnote=\sing@vfootnote
  \ninepoint % reset \baselineskip after leading change
  \message{(Single column)}%
}

\def\singlecoloutput{%
  \shipout\vbox{\PageHead\vbox to \PageHeight{\pagebody\vss}\PageFoot}%
  \advancepageno
  \ifplate@page
    \shipout\vbox{%
      \sp@pagetrue
      \def\sp@type{plate}%
      \global\plate@pagefalse
      \PageHead\vbox to \PageHeight{\unvbox\plt@box\vfil}\PageFoot%
    }%
    \message{[plate]}%
    \advancepageno
  \fi
  \ifnum\outputpenalty>-\@MM \else\dosupereject\fi%
}

\def\ItemSep{\vskip\ItemSepamount\relax}

\def\ItemSepbreak{\par\ifdim\lastskip<\ItemSepamount
  \removelastskip\penalty-200\ItemSep\fi%
}

% Modify plain's \endinsert so that the mn's spacing is used

\let\@@endinsert=\endinsert % save plain's original \endinsert

\def\endinsert{\egroup % finish the \vbox
  \if@mid \dimen@\ht\z@ \advance\dimen@\dp\z@ \advance\dimen@12\p@
    \advance\dimen@\pagetotal \advance\dimen@-\pageshrink
    \ifdim\dimen@>\pagegoal\@midfalse\p@gefalse\fi\fi
  \if@mid \ItemSep\box\z@\ItemSepbreak
  \else\insert\topins{\penalty100 % floating insertion
    \splittopskip\z@skip
    \splitmaxdepth\maxdimen \floatingpenalty\z@
    \ifp@ge \dimen@\dp\z@
    \vbox to\vsize{\unvbox\z@\kern-\dimen@}% depth is zero
    \else \box\z@\nobreak\ItemSep\fi}\fi\endgroup%
}

% Footnotes (only enabled in single column)

\def\gobbleone#1{}
\def\gobbletwo#1#2{}
\let\footnote=\gobbletwo % Gobble footnote's unless enabled by \onecolumn
\let\vfootnote=\gobbleone

\def\sing@footnote#1{\let\@sf\empty % parameter #2 (the text) is read later
  \ifhmode\edef\@sf{\spacefactor\the\spacefactor}\/\fi
  \hbox{$^{\hbox{\eightpoint #1}}$}\@sf\sing@vfootnote{#1}%
}

\def\sing@vfootnote#1{\insert\footins\bgroup\eightpoint\b@ls{9pt}%
  \interlinepenalty\interfootnotelinepenalty
  \splittopskip\ht\strutbox % top baseline for broken footnotes
  \splitmaxdepth\dp\strutbox \floatingpenalty\@MM
  \leftskip\z@skip \rightskip\z@skip \spaceskip\z@skip \xspaceskip\z@skip
  \noindent $^{\scriptstyle\hbox{#1}}$\hskip 4pt%
    \footstrut\futurelet\next\fo@t%
}

% Kill footnote rule
\def\footnoterule{\kern-3\p@ \hrule height \z@ \kern 3\p@}

\skip\footins=19.5pt plus 12pt minus 1pt
\count\footins=1000
\dimen\footins=\maxdimen

% for footnotes in double column: use \note{$\star$}{footnote}
\def\note#1#2{%
  \let\@sf=\empty \ifhmode\edef\@sf{\spacefactor\the\spacefactor}\/\fi
  #1\insert\footins\bgroup
    \eightpoint\b@ls{10pt}\rm
    \interlinepenalty\interfootnotelinepenalty
%    \splittopskip\ht\strutbox % top baseline for broken footnotes
    \splitmaxdepth\dp\strutbox \floatingpenalty\@MM
    \leftskip\z@skip \rightskip\z@skip \spaceskip\z@skip \xspaceskip\z@skip
    \noindent\footstrut #1$\,$#2\strut\par
  \egroup
  \@sf\relax}

% Landscape

\def\landscape{%
  \global\TEMPDIMEN=\PageWidth
  \global\PageWidth=\PageHeight
  \global\PageHeight=\TEMPDIMEN
  \global\let\landscape=\relax%         stop them using \landscape again.
  \onecolumn
  \message{(landscape)}%
  \raggedbottom
}

%-------------------------------------------------------%
%               TWO COLUMN OUTPUT ROUTINE               %
%-------------------------------------------------------%

% Very slight redefinition of the \output routine of mn.tex, to allow footnotes.
\output{%
  \ifLeftCOL
    \global\setbox\LeftBOX=\vbox to \ZoneBSize{\box255\unvbox\ZoneBBOX
      \ifvoid\footins\else
        \vskip\skip\footins\unvbox\footins\fi
    }%
    \global\LeftCOLfalse
    \MakeRightCol
  \else
    \setbox\RightBOX=\vbox to \ZoneBSize{\box255\unvbox\ZoneBBOX
      \ifvoid\footins\else
        \vskip\skip\footins\unvbox\footins\fi
    }%
    \setbox\MidBOX=\hbox{\box\LeftBOX\hskip\ColumnGap\box\RightBOX}%
    \setbox\PageBOX=\vbox to \PageHeight{%
      \UnloadZoneA\box\MidBOX\UnloadZoneC}%
    \shipout\vbox{\PageHead\vbox to \PageHeight{\box\PageBOX\vss}\PageFoot}%
    \advancepageno
    \ifplate@page
      \shipout\vbox{%
        \sp@pagetrue
        \def\sp@type{plate}%
        \global\plate@pagefalse
        \PageHead\vbox to \PageHeight{\unvbox\plt@box\vfil}\PageFoot%
      }%
      \message{[plate]}%
      \advancepageno
    \fi
    \global\LeftCOLtrue
    \CleanStack
    \MakePage
  \fi
}

% Startup message

\Warn{\start@mess}

\newif\ifCUPmtplainloaded % for use in documents
\ifprod@font
  \global\CUPmtplainloadedtrue
\fi

 % so articles can see if a format file has been used.

\catcode `\@=12 % @ signs are non-letters

% \dump

% end of mn.tex

%% file: text-sub.bbl
\beginrefs

\bibitem Anderson~E.~R., Duvall~T.~L., Jefferies~S.~M., 1990,
ApJ, 364, 699

\bibitem Antia H.~M., Basu S., 1994, A\&AS, 107, 421

\bibitem Antia~H.~M., Chitre~S.~M., 1997, ApJ, submitted

\bibitem Bahcall~J.~N., Pinsonneault~M.~H., 1995,
{\rm Rev. Mod. Phys.}, 67, 781 

\bibitem Balmforth~N.~J., 1992, MNRAS, 255, 639

\bibitem Basu~S., Antia~H.~M., 1997, MNRAS, in the press

\bibitem Basu~S.  Christensen-Dalsgaard~J., 1997,
A\&A, submitted

\bibitem Basu~S., Christensen-Dalsgaard~J., Schou~J., Thompson~M.~J.,
Tomczyk~S., 1995, in Hoeksema J.~T., Domingo V., Fleck B., 
Battrick B., eds, Proc. Fourth SOHO Workshop: Helioseismology, Volume 2.
ESTEC, Noordwijk, p.~25

\bibitem Basu~S., Christensen-Dalsgaard~J., Schou~J., Thompson~M.~J.,
Tomczyk~S. 1996a, ApJ, 460, 1064

\bibitem Basu~S., Christensen-Dalsgaard~J., Schou~J., Thompson~M.~J.,
Tomczyk~S. 1996b, Bull. Astron. Soc. India, 24, 147

\bibitem Basu~S., Christensen-Dalsgaard~J., P\'erez Hern\'andez, F.,
Thompson~M.~J., 1996c, MNRAS, 280, 651

\bibitem Chandrasekhar~S., 1964,  ApJ,  139, 664

\bibitem Chaplin~W.~J., Elsworth~Y., Howe~R., Isaak~G.~R.,
McLeod~C.~P., Miller~B.~A., van~der~Raay~H.~B., Wheeler~S.~J.,
New~R., 1996a, Sol. Phys., 168, 1

\bibitem Chaplin~W.~J., Elsworth~Y., Howe~R., Isaak~G.~R.,
McLeod~C.~P., Miller~B.~A., New~R., 1996b, MNRAS, 282, L15

\bibitem Christensen-Dalsgaard~J., 1986,
in Gough~D.~O., ed,
Seismology of the Sun and the distant Stars.
Reidel, Dordrecht, p.~23.

\bibitem Christensen-Dalsgaard~J., Berthomieu~G., 1991,
in Cox~A.~N., Livingston~W.~C., Matthews~M., eds,
Solar Interior and Atmosphere.
University of Arizona Press, Tucson, p.~401

\bibitem Christensen-Dalsgaard~J., Gough~D.~O.  P\'erez Hern\'andez~F., 1988,
MNRAS, 235, 875

\bibitem Christensen-Dalsgaard~J., Gough~D.~O., Thompson~M.~J., 1991,
ApJ, 378, 413

\bibitem Christensen-Dalsgaard~J., D\"appen~W., Ajukov~S.~V., et al., 1996,
Science, 272, 1286

\bibitem Cox~A.~N., Kidman~R.~B., 1984, in
Theoretical problems in stellar stability and oscillations.
Institut d'Astrophysique, Li\`ege, p.~259

\bibitem D\"appen~W., Gough~D.~O., Kosovichev~A.~G., Thompson~M.~J., 1991,
in Gough~D.~O., Toomre~J., eds,
Lecture Notes in Physics,  388. Springer, Heidelberg, p.~111

\bibitem  Dziembowski~W.~A., Pamyatnykh~A.~A., Sienkiewicz~R., 1990,
\hfill\break
MNRAS, 244, 542

\bibitem Dziembowski~W.~A., Pamyatnykh~A.~A., Sienkiewicz~R., 1991,
\hfill\break
MNRAS,  249, 602

\bibitem Dziembowski~W.~A., Goode~P.~R., Pamyatnykh~A.~A., Sien\-kiewich~R.,
1994,  ApJ,  432, 417


\bibitem Elsworth~Y., Howe~R., Isaak~G.~R., McLeod~C.~P., 
Miller~B.~A., New~R., Speake~C.~C., Wheeler~S.~J., 1994, ApJ, 434, 801

\bibitem Gough D. O., 1990, in Osaki Y., Shibahashi H., eds, Lecture
Notes in Physics, 367. Springer, Berlin, p.~283

\bibitem Gough~D.~O., Kosovichev~A.~G., 1993,  MNRAS,  264, 522

\bibitem Gough~D.~O., Kosovichev~A.~G., Toutain~T., 1995,  Solar Phys.,
 157, 1

\bibitem Gough~D.~O., Kosovichev~A.~G., Toomre~J., et al., 1996,  Science,
 272, 1296

\bibitem Grevesse, N., Noels, A. 1993, in
Prantzos N., Vangioni-Flam E., Cass\'e M., eds,
Origin and evolution of the Elements.
Cambridge Univ. Press, Cambridge, p.~15

\bibitem Guenther~D.~B., 1989, ApJ, 339, 1156 

%\bibitem Iglesias~C.~A., Rogers~F.~J., 1996, ApJ, 464, 943

\bibitem Iglesias~C.~A., Rogers~F.~J., Wilson~B.~G., 1992, ApJ, 397, 717

\bibitem Kosovichev~A.~G., Christensen-Dalsgaard~J., D\"appen, W.,
Dziembowski~W.~A., Gough~D.~O., Thompson~M.~J., 1992,
 MNRAS,  259, 536

\bibitem Kurucz~ R. L.,  1991, in
Crivelli~L., Hubeny I., Hummer D.~G., eds, 
Stellar atmospheres: beyond classical models.
NATO ASI Series, Kluwer, Dordrecht, p.~441

%\bibitem Libbrecht~K.~G., Woodard~M.~F., 1991, Science, 253, 152

\bibitem Libbrecht~K.~G., Woodard~M.~F., 1990, Nature, 345, 779

\bibitem Mihalas~D., D\"appen~W., Hummer~D.~G., 1988,
ApJ, 331, 815 

\bibitem
Pijpers~F.~P., Thompson~M.~J.,  1992,  A\&A,  262, L33

\bibitem Rogers~F.~J., Swenson~F.~J.,  Iglesias~C.~A., 1996, ApJ, 456, 902

\bibitem Schou~J., 1992.
Ph.D thesis, Aarhus University.

\bibitem Tomczyk~S., Streander~K., Card~G., Elmore~D., Hull~H.,
Cacciani~A., 1995, Sol. Phys., 159, 1

\bibitem Toutain~T.,  Appourchaux~T., 1994, A\&A, 289, 649

\bibitem Unno~W., Osaki~Y., Shibahashi~H., 1989, Non-radial Oscillations
of Stars, 2nd ed.. Univ. of Tokyo Press, Tokyo


\endrefs
